\documentclass[iop]{emulateapj}

\pdfoutput=1

\newcommand\xone{x_1}
\newcommand\xtwo{x_2}
\newcommand\Mgas{M_{x_1}}
\newcommand\cs{c_s}
\newcommand\aSig{{\overline \Sigma}_{\rm cl}}
\newcommand\bSig{\langle\Sigma\rangle}
\newcommand\SigSFR{\Sigma_{\rm SFR}}
\newcommand\mSFR{{\rm SFR_{\rm max}}}
\newcommand\tSig{\Sigma_{\rm th}}
\newcommand\mcp{\dot M_{*,\rm CP}}
\newcommand\Mcl{M_{\rm cloud}}
\newcommand\Min{\dot M_{\rm NR}}
\newcommand\Mring{M_{\rm NR}}
\newcommand\Rring{r_{\rm NR}}

\newcommand\SFR{{\rm SFR}}
\newcommand\dsf{\Delta t_{\rm SF}}
\newcommand\Mtotal{M_*^{\rm tot}}
\newcommand\Mstar{M_*}
\newcommand\MBH{M_{\rm BH}}
\newcommand\NSN{{\mathcal N}_{\rm SN}}

\newcommand\fmom{f_{\rm mom}}
\newcommand\rsh{R_{\rm sh}}
\newcommand\rSF{R_{\rm SF}}
\newcommand\Msun{\; {\rm M}_{\odot}}
\newcommand\kms{\; {\rm km}\;{\rm s}^{-1}}
\newcommand\pc{\;{\rm pc}}
\newcommand\erg{\;{\rm erg}}
\newcommand\kpc{\;{\rm kpc}}
\newcommand\freq{\kms\kpc^{-1}}

\newcommand\yr{\; {\rm yr}}
\newcommand\Myr{\;{\rm Myr}}
\newcommand\Gyr{\;{\rm Gyr}}
\newcommand\Aunit{\Msun \yr^{-1}}
\newcommand\SFRunit{\Msun \yr^{-1}\kpc^{-2}}
\newcommand\Surf{\Msun\;{\rm pc^{-2}}}
\newcommand\torb{t_{\rm orb}}
\newcommand\tbar{\tau_{\rm bar}}
\newcommand\simgt{\lower.5ex\hbox{$\; \buildrel > \over \sim \;$}}
\newcommand\simlt{\lower.5ex\hbox{$\; \buildrel < \over \sim \;$}}

\shorttitle{Star Formation in Barred Galaxies}
\shortauthors{Seo \& Kim}
\begin{document}

\title{Star Formation in Nuclear Rings of Barred Galaxies}
\author{Woo-Young Seo and  Woong-Tae Kim} %%
\affil{Center for the Exploration of the Origin of the Universe
(CEOU), Astronomy Program, Department of Physics \& Astronomy, Seoul
National University, \\ Seoul 151-742, Republic of Korea}%%
\affil{FPRD, Department of Physics
\& Astronomy, Seoul National University, Seoul 151-742, Republic of
Korea}%%

\email{seowy@astro.snu.ac.kr,  wkim@astro.snu.ac.kr}
\slugcomment{Accepted for publication in the ApJ}

\begin{abstract}
Nuclear rings in barred galaxies are sites of active star formation.
We use hydrodynamic simulations to study temporal and spatial
behavior of star formation occurring in nuclear rings of barred
galaxies where radial gas inflows are triggered solely by a bar
potential. The star formation recipes include a density threshold,
an efficiency, conversion of gas to star particles, and delayed
momentum feedback via supernova explosions. We find that star
formation rate (SFR) in a nuclear ring is roughly equal to the mass
inflow rate to the ring, while it has a weak dependence on the total
gas mass in the ring. The SFR typically exhibits a strong primary
burst followed by weak secondary bursts before declining to very
small values. The primary burst is associated with the rapid gas
infall to the ring due to the bar growth, while the secondary bursts
are caused by re-infall of the ejected gas from the primary burst.
While star formation in observed rings persists episodically over a
few Gyr, the duration of active star formation in our models lasts
for only about a half of the bar growth time, suggesting that the
bar potential alone is unlikely responsible for gas supply to the
rings. When the SFR is low, most star formation occurs at the
contact points between the ring and the dust lanes, leading to an
azimuthal age gradient of young star clusters.  When the SFR is
large, on the other hand, star formation is randomly distributed
over the whole circumference of the ring, resulting in no apparent
azimuthal age gradient. Since the ring shrinks in size with time,
star clusters also exhibit a radial age gradient, with younger
clusters found closer to the ring.  The cluster mass function is
well described by a power law, with a slope depending on the SFR.
Giant gas clouds in the rings have supersonic internal velocity
dispersions and are gravitationally bound.
\end{abstract}

\section{Introduction\label{sec:intro}}

Nuclear rings in barred galaxies are sites of intense star formation
(e.g.,
\citealt{bur60,san61,phi96,but96,kna06,maz08,com10,san10,maz11,hsi11}).
These rings are thought to form as a result of nonlinear
interactions of gas with a non-axisymmetric bar potential (e.g.,
\citealt{com85,shl90,ath92,hel94,kna95,but96,com01,pin95,reg03}).
Due to the bar torque, the gas readily forms dust-lane shocks in the
bar region and flows inward along the dust lanes. The inflowing gas
speeds up gradually in the azimuthal direction as it moves inward,
and shapes into a ring very close to the galaxy center (e.g.,
\citealt{kim12}). Consequently, nuclear rings have very large
surface densities and short dynamical time scales, capable of
triggering starburst activity.

There are some important observational results that may provide
clues as to how star formation occurs in the nuclear rings. First of
all, observations indicate that the star formation rate (SFR) in the
nuclear rings appears to vary with time, and differs considerably
from galaxy to galaxy. Analyses of various population synthesis
models for a sample of galaxies reveal that the strength of observed
emission lines from the nuclear rings is best described by multiple
starburst activities over the last $0.5\Gyr$ or so, rather than by a
constant star formation rate (e.g., \citealt{all06,sar07}). For 22
nuclear rings, \citet{maz08} found that the present SFRs are widely
distributed in the range $0.1$--$10\Aunit$. It appears that the SFR
is largely insensitive to the total molecular mass in a ring, but
can be strongly affected by the bar strength. For instance, the ring
in a strongly-barred galaxy NGC 4314 has a relatively low SFR at
$\sim0.1\Aunit$ \citep{ben02}, which is about an order of magnitude
smaller than that in a weakly-barred galaxy NGC 1326 \citep{but00},
although the total molecular mass contained in the ring is within a
factor of two. In fact, the SFRs given in \citet{maz08} combined
with the bar strength presented by \citet{com10} show that
strongly-barred galaxies tend to have a very small SFR in the rings,
while weakly-barred galaxies have a wide range of the SFRs.

Second, based on the spatial distributions of star-forming regions
in nuclear rings, \citet{bok08} proposed two models of star
formation: ``popcorn'' and ``pearls on a string'' models (see also
\citealt{san10}). In the first popcorn model,  star formation occurs
in dense clumps that are randomly distributed along a nuclear ring.
This type of star formation, presumably caused by gravitational
instability of the ring itself \citep{elm94}, does not produce a
systematic gradient in the ages of young star clusters along the
azimuthal direction (see also, e.g., \citealt{ben02,bra12}). In the
second pearls-on-a-string model, on the other hand, star formation
takes place preferentially at the contact points between a ring and
dust lanes.  This may happen because gas clouds with the largest
densities are usually placed at the contact points due to orbit
crowding (e.g., \citealt{ken92,rey97,koh99,hsi11}). Since star
clusters age as they orbit along the ring, this model naturally
predicts a bipolar azimuthal age gradient of star clusters starting
from the contact points (see also, e.g.,
\citealt{ryd01,all06,maz08,bok08,ryd10,van13}). \citet{maz08} found
that $\sim50\%$ of the nuclear rings in their sample galaxies show
azimuthal age gradients and that such galaxies have, on average, a
larger value of the mean SFR than those without noticeable age
gradients.

Another interesting observational result concerns radial locations
of star clusters relative to the nuclear rings. In some galaxies
such as NGC 1512 \citep{mao01} and NGC 4314 \citep{ben02}, young
star clusters are located at larger radii than the dense gas of
nuclear rings. \citet{mar03} also reported that out of 123 barred
galaxies in their sample, eight galaxies have strong nuclear
spirals, all of which have star-forming regions outside the rings.
By analyzing multi-waveband HST archive data of NGC 1672,
\citet{jan13} recently identified hundreds of young and old star
clusters with ages in the range $\sim 1-10^3\Myr$. They found that
the clusters in the nuclear regions exhibit a systematic positive
radial age gradient, such that older clusters tend to be located at
larger galactocentric radii, farther away from the ring. Proposed
mechanisms for the radial age gradient include the decrease in the
ring size due to angular momentum loss \citep{reg03} and migration
of clusters due to tidal interactions with the ring \citep{van09}.

Numerical simulations have been a powerful tool to study formation
and evolution of bar substructures such as dust lanes, nuclear
rings, and nuclear spirals (e.g.,
\citealt{san76,ath92,pin95,eng97,pat00,mac02,reg03,reg04,
ann05,tha09,kim12,ksk12}). In particular, \citet{ath92} showed that
dust lanes are shocks formed at the downstream side from the bar
major axis. Dust lanes tend to be shorter and located closer to the
bar major axis as the gas sound speed increases
\citep{eng97,pat00,kim12}.

Very recently, \citet[hereafter Paper I]{ksk12} ran various models
with differing bar strength and demonstrated that nuclear rings form
not by resonant interactions of the gas with the bar potential, as
was previously thought, but instead by the centrifugal barrier that
the inflowing gas with non-vanishing angular momentum cannot
overcome. According to this idea, a more massive bar forms stronger
dust-lane shocks which remove angular momentum more efficiently from
the gas, so that the inflowing gas is able to move inward closer to
the galaxy center,  forming a smaller nuclear ring. This turns out
entirely consistent with the observational result of \citet{com10}
that ``stronger bars host smaller rings''. Magnetic stress at the
dust lanes takes away angular momentum additionally, leading to an
even smaller ring compared to the unmagnetized counterpart
\citep{ks12}. Paper I also showed that nuclear spirals that form
inside nuclear rings unwind with time due to the nonlinear effect
\citep{lee99}, with an unwinding rate higher for a stronger bar.
Thus, the probability of having more tightly wound spirals is larger
for galaxies with a weaker bar, consistent with the observational
result of \citet{pee06}.

While the numerical studies mentioned above are useful to understand
gas dynamics in the central regions of barred galaxies, they are
without self-gravity and/or prescriptions for star formation. There
have been only a few numerical studies that considered star
formation in nuclear rings in a self-consistent way. \citet{hel94}
studied star formation in galactic disks that are unstable to bar
formation. Using a smoothed particle hydrodynamics (SPH) combined
with $N$-body method, they found that star formation in barred
galaxies occurs episodically, with a time scale of $\sim10\Myr$, and
that the associated SFR is well correlated with the mass accretion
rate to the central black hole (BH). These were confirmed by
\citet{kna95} who also found that turbulence driven by star
formation tends to widen nuclear rings. \citet{fri95} used another
SPH$+N$-body method to run various models with differing parameters,
finding that star formation in the nuclear regions first experiences
a burst phase before entering a quiescent phase (see also
\citealt{mar97}). Since these authors employed a small number ($\sim
10^4$) of gas particles in their models, however, they were unable
to resolve the nuclear regions well. \citet{ksj11} ran SPH
simulations for star formation specific to the central molecular
zone in the Milky Way. While \citet{dob10} studied cluster age
distributions in spiral and barred galaxies, their results were
based on SPH simulations that did not consider star formation and
feedback.

In this paper, we extend Paper I by including self-gravity and a
prescription for star formation feedback. We focus on temporal and
spatial distributions of star formation occurring in nuclear rings
of strongly-barred galaxies. Unlike the previous SPH simulations
with star formation, our models use a grid-based, cylindrical code
with high spatial resolution in the central regions. We also allow
for time delays between star formation and feedback, which is
crucial to study age gradients of star clusters that form in nuclear
rings. Our main objectives are to address important questions such
as what controls the SFR in the nuclear rings and what are
responsible for the presence (or absence) of the age gradients of
star clusters in the rings, mentioned above.

We take a simple galaxy model in which a self-gravitating gaseous
disk with either uniform or exponential density distribution is
placed under the influence of a non-axisymmetric bar potential. We
implement a stochastic prescription for star formation that takes
allowance for a threshold density as well as a star formation
efficiency. Star formation feedback is treated only through direct
momentum injections from supernova (SN) explosions occurring
$10\Myr$ after star formation events.  By considering an isothermal
equation of state, we do not consider gas cooling and heating, and
radiative feedback, which may be important in regulating star
formation in disk galaxies (e.g.,
\citealt{oml10,ost11,kko11,she12}). In our models, the bar potential
is turned on slowly over time, which not only represents a situation
where the bar forms and grows but also helps avoid abrupt gas
responses. In each model, we measure the SFR in the ring and study
its dependence on various quantities such as the gas mass in the
ring, mass inflow rates to the central regions, bar growth time,
etc. We also explore temporal and spatial variations of star-forming
regions and their connection to the SFR. In addition, we study
physical properties of star clusters and gas clouds in the rings and
compare them with observational results available.

We remark on a few important limitations of our models from the
outset. First of all, our gaseous disks are two-dimensional and
razor-thin. This ignores potential dynamical consequences of
vertical gas motions and related mixing, which was shown important
in inducing non-steady gas motions across spiral shocks (e.g.,
\citealt{kim06,kko06,kko10}). Second, we adopt an isothermal
equation of state for the gas, corresponding to the warm phase, and
do not consider radiative cooling and heating required for
production and transitions of multiphase gas (e.g.,
\citealt{fie65,wol03,mck07}). We also ignore the effects of
outflows, winds, and radiative feedback from young stars, which may
be of crucial importance in setting up the equilibrium pressure in
galactic planes, thereby regulating star formation in disk galaxies
(e.g., \citealt{oml10,ost11,she12}). Finally, we in the present work
do not consider the effects of spiral arms that may supply gas to
the bar regions. With these caveats, the numerical models presented
in this paper should be considered as a first step toward more
realistic modeling of star formation in nuclear rings.

This paper is organized as follows. In Section 2, we describe the
numerical methods and parameters used for our time-dependent
simulations.  In Section 3, we present the temporal evolution of
SFRs, and their dependence on the model parameters. The age
gradients of star clusters in the azimuthal and radial directions as
well as properties of star clusters and dense clouds are discussed
in Section 4. In Section 5, we summarize our main results and
discuss their astronomical implications.

\section{Model and Method}\label{sec:method}

To study star formation in nuclear rings of barred galaxies, we
extend the numerical models studied in Paper I by including
self-gravity, conversion of gas to stars, and feedback from star
formation.  In this section, we briefly summarize the current models
and describe our handling of star formation and feedback. The reader
is referred to Paper I for more detailed description of the
numerical models.

\subsection{Galaxy Model}\label{sec:model}

We initially consider an infinitesimally-thin, rotating disk. The
disk is assumed to be unmagnetized and isothermal with sound speed
of $\cs=10\kms$. The external gravitational potential responsible
for the disk rotation consists of four components: a stellar disk, a
stellar bulge, a non-axisymmetric stellar bar, and a central BH with
mass $\MBH=4\times10^7\Msun$. This gives rise to a rotation curve
that is almost flat at $v_c \sim200\kms$ in the bar region and its
outside. The presence of the BH makes the rotation velocity rise as
$v_c\propto (\MBH/r)^{1/2}$ toward the galaxy center. The bar
potential is modeled by a \citet{fer87} prolate spheroid with
semi-major and minor axes of $5\kpc$ and $2\kpc$, respectively.  The
bar is rigidly rotating with a pattern speed $\Omega_b=33\freq$,
which places the corotation resonance radius at $r=6\kpc$ and the
inner Lindblad resonance (ILR) radius at $r=2.2\kpc$. The
corresponding orbital time is $\torb=2\pi/\Omega=186\Myr$.  In our
models, the bar potential is turned on over the bar growth time
scale $\tbar$, while the central density of the spheroidal component
(bar plus bulge) is kept fixed. We vary $\tbar$ to study situations
where the bar grows at a different rate. The mass of the bar, when
it is fully turned on, is set to 30\% of the total mass of the
spheroidal component within $10\kpc$. All the models are run until
$1\Gyr$.

%table1
\begin{deluxetable}{lccc}
\tabletypesize{\footnotesize} \tablewidth{0pt} \tablecaption{Model
Parameters \label{tbl:model}} \tablehead{ \colhead{Model}   &
\colhead{$\Sigma_0 (\Msun\rm\;pc^{-2})$} & \colhead{$\tbar/\torb$}
&
\colhead{$\fmom$}            \\
\colhead{(1)} & \colhead{(2)} & \colhead{(3)} & \colhead{(4)} }
\startdata
noSG  &  20 & 1 & 0.\\
U05   &  5  & 1 & 0.75 \\
U10   &  10 & 1 & 0.75 \\
U20   &  20 & 1 & 0.75 \\
U30   &  30 & 1 & 0.75 \\
\hline
M25   &  20 & 1 & 0.25 \\
M50   &  20 & 1 & 0.50 \\
\hline
FB05  &  20 & 0.5 & 0.75 \\
FB20  &  20 & 2   & 0.75 \\
FB40  &  20 & 4   & 0.75 \\
\hline
E30   & 30 & 1  & 0.75  \\
E50   & 50 & 1  & 0.75  \\
E100  & 100& 1  & 0.75
\enddata
\tablecomments{Gas surface density in Models with the prefix ``E''
initially have an exponential distribution $\Sigma=\Sigma_0
\exp{(-r/3.5\kpc)}$. All the other models have a uniform density
distribution $\Sigma=\Sigma_0$.}
\end{deluxetable}

As in Paper I, we integrate the basic equations of ideal
hydrodynamics in a frame corotating with the bar. We use the CMHOG
code in cylindrical polar coordinates $(r, \phi)$. CMHOG is
third-order accurate in space and has very little numerical
diffusion \citep{pin95}. To resolve the central region with high
accuracy, we set up a logarithmically-spaced cylindrical grid over
$r=0.05\kpc$ to $8\kpc$. The number of zones in our models is 1024
in the radial direction and 632 in the azimuthal direction covering
the half-plane from $\phi=-\pi/2$ to $\pi/2$. The corresponding
spatial resolution is $0.25\pc$, $5\pc$, and $40\pc$ at the inner
boundary, at $r=1\kpc$ where most star formation takes places, and
at the outer radial boundary, respectively. We adopt the outflow and
continuous boundary conditions at the inner and outer radial
boundaries, respectively, while taking the periodic boundary
conditions at $\phi=\pm \pi/2$.

Our models consider conversion of gas to stars, as will be explained
in Section \ref{sec:sf} in detail. Since the total gas mass
transformed to stars is significant, it is important to evolve them
under the combined gravitational potential of the gas and stars. At
each time step, we calculate the stellar surface density on the grid
points via the triangular-shaped-cloud assignment scheme
\citep{hoc88} from the distribution of stellar particles. We then
solve the Poisson equation to obtain the gravitational potential of
the total (gas plus star) surface density, using the \citet{kal71}
method presented in \citet{she08}.\footnote{We ignore the gravity
from the initial gas distribution in order to make the initial
rotation curve the same with that in the non-self-gravitating
counterpart.} To allow for dilution of gravity due to finite
thickness $H$ of the combined disk, we take $H/r=0.1$ as a softening
parameter in the potential calculation.

To explore the dependence of SFR upon the total gas content and the
way the gas is spatially distributed, we initially consider gaseous
disks with either uniform surface density $\Sigma_{0}$ or an
exponential distribution $\Sigma_{0} \exp(-r/R_d)$ with the scale
length of $R_d=3.5\kpc$.   We also vary the fraction $\fmom$ of the
radial momentum from SNe imparted to the disk in the in-plane
direction relative to what would be the total radial momentum in a
three-dimensional uniform medium (see Section \ref{sec:sf}). We run
a total of 13 models that differ in $\Sigma_0$, $\tbar$, and
$\fmom$. Table \ref{tbl:model} lists the model parameters. Column
(1) lists each model.  Models with the prefix ``E" have an
exponential disk, while all the others have a uniform disk. Model
noSG is a control model that does not include self-gravity and star
formation. Column (2) lists $\Sigma_0$ of the disk.  Column (3)
gives the bar growth time $\tbar$ in units of $\torb$, while Column
(4) gives $\fmom$. We take Model U20 with $\Sigma_0=20\Surf$,
$\tbar/\torb=1$, and $\fmom=0.75$ as our fiducial model.  All the
models initially have a Toomre $Q$ parameter greater than unity, so
that they are gravitationally stable in the absence of a bar
potential. However, nuclear rings that form near the center achieve
large density, enough to undergo runaway collapse to form stars.

\subsection{Star Formation and Feedback}\label{sec:sf}

To model star formation and ensuing feedback, we first identify
high-density regions whose average surface density $\bSig$ within a
radius $\rSF$ exceeds a critical density.  The natural choice for
the threshold density would be
\begin{equation}\label{eq:th}
\tSig=\frac{\cs^2}{2G\rSF} = 1160\Surf\left(\frac{\cs}{10\kms}\right)^2
\left(\frac{\rSF}{10\pc}\right)^{-1},
\end{equation}
from the Jeans condition.  While it is desirable to choose a small
value for the sizes of star-forming regions, we take $\rSF=10\pc$
because of numerical resolution: a star-forming cloud at
$r\sim1\kpc$ encloses typically $\sim13$ grid points.

Not all clouds with $\bSig\geq \tSig$ immediately undergo
gravitational collapse and star formation since we need to consider
the star formation efficiency as well as the computational time step
\citep{kko11}. The SFR expected from a cloud with mass $\Mcl=\pi
\rSF^2\bSig$, from the \citet{sch59} law, is
\begin{equation}
\SFR = \epsilon_{\rm ff}\frac{\Mcl}{t_{\rm ff}} \;\;\;{\rm for}\;\;
\bSig \geq \tSig,
\end{equation}
where $\epsilon_{\rm ff}$ is the star formation efficiency per
free-fall time, $t_{\rm ff}$, defined by
\begin{equation}
t_{\rm ff} = \left( \frac{3\pi}{32G\langle\rho\rangle}\right)^{1/2}
= 3.4 \Myr \left(\frac{\bSig}{1160\Surf}\right)^{-1/2},
%=  \frac{1.14\times10^8}{\sqrt{\Sigma}}\yr.
\end{equation}
assuming a disk scale height of
$100\pc$. We take $\epsilon_{\rm ff} = 0.01$, consistent with
theoretical and observational estimates (e.g.,
\citealt{kru05,kru07}).

The star formation probability of an eligible cloud with $\bSig\geq
\tSig$ in a time interval $\Delta t$ is then given by $p=1-
\exp(-\epsilon_{\rm ff}\Delta t/t_{\rm ff}) \approx \epsilon_{\rm
ff}\Delta t/t_{\rm ff}$ (e.g., \citealt{hop11}).  For a given
computational time step $\Delta t$, the probability $p$ calculated
in our models is typically $\sim 10^{-6}-10^{-5}$, much smaller than
unity. In each time step, we thus generate a uniform random number
$\mathcal{N} \in [0,1)$, and turn on star formation only if
$\mathcal{N} < p$.   When a cloud undergoes star formation, we
create a particle with mass $\Mstar$, and convert 90\% of the cloud
mass to the particle mass. The initial position and velocity of the
particle are set equal to the density-weighted mean values of the
parent cloud within $\rSF$. Each particle has a mass in the range
$\Mstar\sim10^5-10^7\Msun$, which is about $\sim 1-10^2$ times
larger than the masses of observed clusters in nuclear rings (e.g.,
\citealt{mao01,ben02}; see also \citealt{por10}). Therefore, a
massive single particle in our models can be regarded as
representing an unresolved group of star clusters rather than an
individual cluster.

We treat SN feedback using simple momentum input to the surrounding
gaseous medium.  We consider only Type II SN events since our models
run only until $1\Gyr$.  Since we do not resolve individual stars in
a cluster or their group, we assume that all SN explosions occur
simultaneously. Stars with mass between $8\Msun$ and $40\Msun$
explode as Type II SNe \citep{heg03}, which comprise about 7\% of
the cluster mass under the \citet{kro01} initial mass function. The
mean mass of SN progenitors is then $\sim 14\Msun$, indicating that
the number of SNe exploding from a cluster (or their group) with
mass $\Mstar$ is $\NSN = \Mstar/(200\Msun)$, with the total ejected
mass $M_{\rm ejecta} = 0.07\Mstar$ returning back to the ISM.
Between star formation and SN explosions, we allow a time delay of
$10\Myr$, corresponding to the mean life time of Type II SN
progenitors (e.g., \citealt{lej01}). Note that the orbital time of
gas in nuclear rings is typically $\sim25\Myr$ in our models, so
that star clusters move by about $150^\circ$ in the azimuthal angle
from the formation sites before experiencing SN explosions.

Each feedback, corresponding to $\NSN$ simultaneous SNe, injects
mass and radial momentum in the form of an expanding shell. In the
momentum-conserving stage, a single SN with energy $10^{51}\erg$
would drive radial momentum $P_{\rm rad,3D} = 3\times10^5
\epsilon_0^{7/8} (\Sigma/1\Surf)^{-1/4}\Msun \kms$ to the
surrounding gas if the background medium is uniform and in three
dimensions, where $\epsilon_0$ is the SN energy in units of
$10^{51}\erg$ (e.g., \citealt{che74,shu80,cio88}). The dependence of
$P_{\rm rad,3D}$ on $\Sigma$ is due to the fact that the shell
expansion in the Sedov phase is slow when the background surface
density is large. Since our model disks are razor-thin by ignoring
the vertical direction, the momentum imparted to the gas in the
simulation domain would be smaller than $P_{\rm rad,3D}$.  Let
$\fmom$ denote the fraction of the total radial momentum that goes
into the in-plane direction. If the expansion is isotropic,
$\fmom\sim75\%$ \citep{kko11}, but $\fmom$ can be smaller in  a
vertically stratified disk since it is easier for a shell to expand
along the vertical direction. The total momentum of a shell from
each feedback is thus set to
\begin{equation}
P_{\rm sh} = 3\times10^5  \fmom \; \NSN^{7/8}
\left(\frac{\Sigma}{1\Surf}\right)^{-1/4}\Msun \kms.
\end{equation}
In this paper, we take $\fmom=0.75$ as a standard value, but run
some models with lower $\fmom$ to study the effect of $\fmom$
on the SFR (see Table \ref{tbl:model}).

As the initial radius of a shell, we take $\rsh=40\pc$,
corresponding to the shell size at the end of the Sedov phase when
$\NSN=10^3$ and the background density is $\Sigma=10^3 \Surf$, the
typical mean density of nuclear rings when star formation is active.
When feedback occurs from a particle, we redistribute the mass and
momentum within a circular region with radius $\rsh$ centered at the
particle by taking their spatial averages. We then add the shell
momentum density
\begin{equation}\label{eq:mom}
\Sigma {\bf v}_{\rm sh} = \left\{\begin{array}{ll}
     P_{\rm max} \left(\frac{R}{\rsh^2}\right)\mathbf{R}, & r\leq \rsh,\\
     0 , & r>\rsh,
\end{array}\right.
\end{equation}
to the gas momentum density in the in-plane direction, and $M_{\rm
ejecta}/(\pi\rsh^2)$ to the gas surface density, while reducing the
particle mass by $M_{\rm ejecta}$. In equation (\ref{eq:mom}),
$\mathbf{R}$ denotes the position vector relative to the particle
location and $P_{\rm max} = 2P_{\rm sh}/(\pi \rsh^2)$ is the
momentum per unit area at $r=\rsh$. Note that $v_{\rm sh}(R)\propto
R^2$ ensures an initially divergence-free condition at the feedback
center (e.g., \citealt{kko11}).

\section{Star Formation in Nuclear Rings}\label{sec:res}

In this section, we first describe overall evolution of our
numerical simulations. We then present the temporal variations of
SFRs and their dependence on the gas mass, the bar growth time, and
$\fmom$, as well as the relationship between the SFR surface density
and the gas surface density. The properties of star clusters and gas
clouds in the rings are analyzed in the next section.

%fig1
\begin{figure*}
\hspace{0.5cm}\includegraphics[angle=0, width=17cm]{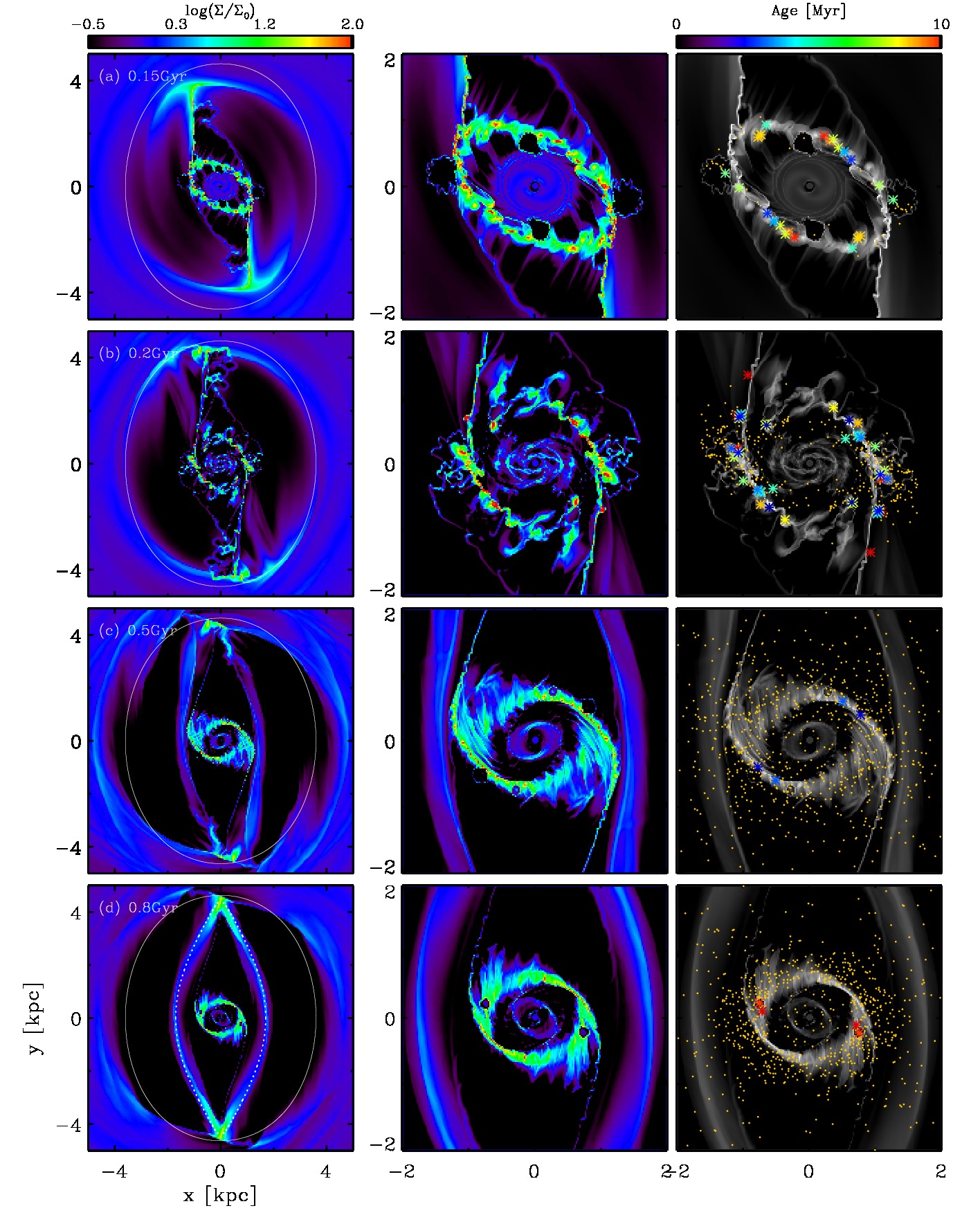}
\caption{Snapshots of logarithm of gas density (color scale) as well
as the locations of star clusters in Model U20 at $t=0.15, 0.2, 0.5,
0.8\Gyr$. The left panels show the $5\kpc$ regions, while the middle
and right panels zoom in the central $2\kpc$ regions. The upper left
colorbar labels $\log (\Sigma/\Sigma_0)$. The solid ovals in the
left panels draw the outermost $\xone$-orbit that cuts the $x$- and
$y-$axes at $x_c=3.6\kpc$ and $y_c=4.7\kpc$. The dotted curve in (d)
is an $\xone$-orbit with $x_c=1.7\kpc$ and $y_c=4.4\kpc$ that traces
the inner ring. Small dots in the right panels denote clusters older
than $10\Myr$, while asterisks represent clusters younger than
$10\Myr$, with the upper right colorbar displaying their ages.
\label{fig:snap}}
\end{figure*}

%fig2
\begin{figure}
\hspace{0.5cm}\includegraphics[angle=0, width=8cm]{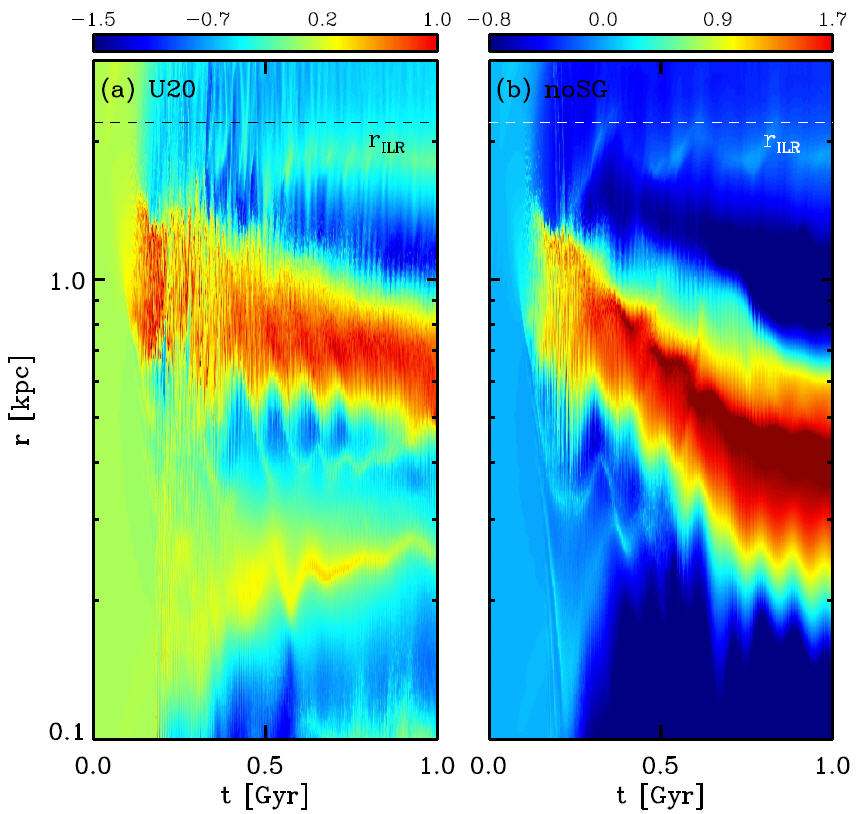}
\caption{Temporal and radial variations of the azimuthally averaged
surface density in logarithmic scale for (a) Model U20 and (b) Model
noSG. The horizontal dashed line in each panel marks the location of
the inner Lindblad resonance. The colorbars label
$\log(\Sigma/\Sigma_0)$. The decreasing rate of the ring size is
smaller when self-gravity and star formation are included.
\label{fig:ring}}
\end{figure}

\subsection{Overall Gas Evolution}\label{sec:over}

We begin by presenting evolution of our standard model U20 that has
a uniform density $\Sigma_0=20\Surf$ initially.  Evolution of the
other models is qualitatively similar, although more massive disks
show more active star formation. In all models, star formation
occurs mostly in the nuclear rings.\footnote{In Model U30, the
bar-end regions, often called ansae, form stars as well, although
the associated SFR is less than 3\% compared to the ring star
formation.} Figure \ref{fig:snap} plots snapshots of the gas
distributions as well as the positions of star clusters in Model U20
at a few selected epochs. The left panels show the gas density in
logarithmic scales in the $5\kpc$ regions, with the solid ovals
indicating the outermost $\xone$-orbit that cuts the $x$- and
$y$-axes at $x_c=3.6\kpc$ and $y_c=4.7\kpc$, respectively. The
middle and right panels zoom in the central $2\kpc$ regions. In the
right panels, small dots indicate star clusters older than $10\Myr$,
while asterisks correspond to young clusters with age $<10\Myr$: the
upper right colorbar represents their ages. In all panels, the bar
is oriented vertically along the $y$-axis and remains stationary.
The gas inside the corotation resonance is rotating in the
counterclockwise direction.

%fig3
\begin{figure*}
\hspace{0.5cm}\includegraphics[angle=0, width=17cm]{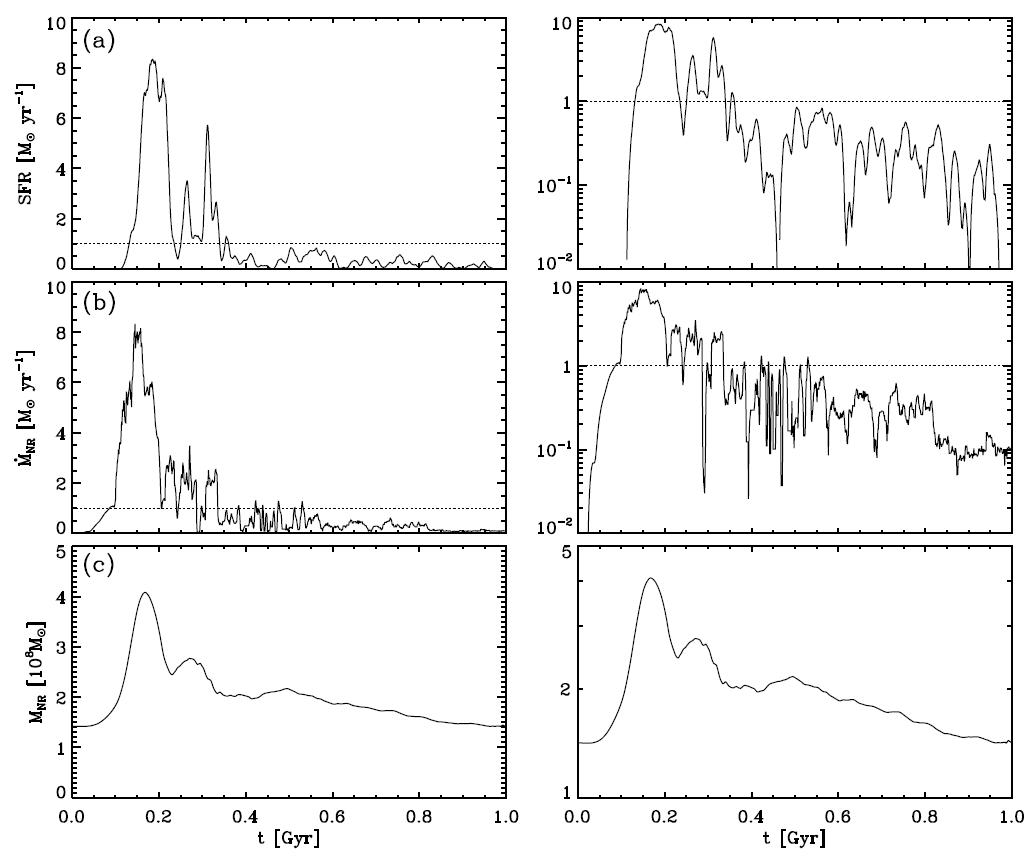}
\caption{Temporal variations of (a) the SFR, (b) the mass inflow
rate $\Min$ to the nuclear ring, and (c) the total gas mass $\Mring$
in the nuclear ring of Model U20. The ordinate is in linear scale in
the left panels, while it is in logarithmic scale in the right
panels.  In (a) and (b), the horizontal dotted lines indicate the
reference rate of $1\Aunit$. \label{fig:U20}}
\end{figure*}

Early evolution of Model U20 before $t=0.1\Gyr$ is not much
different from non-self-gravitating models presented in
\citet{kim12}. The imposed non-axisymmetric bar potential perturbs
gas orbits to create overdense ridges at the far downstream side
from the bar major axis. Only the region inside the outermost
$\xone$-orbit responds strongly to the bar potential. Before the bar
is fully turned on (i.e., $t<\tbar$), no closed $\xone$- and
$\xtwo$-orbits exist since the external gravitational potential
varies with time. During this time, the gas streamlines are highly
transient as they try to adjust to the time-varying potential. The
overdense ridges grow with time and move slowly toward the bar major
axis, developing into dust-lane shocks. The gas passing through the
shocks loses angular momentum and flows radially inward along the
dust lanes. The radial velocity of the inflowing gas is so large
that it is not halted at the ILR (Paper I). It gradually rotates
faster due to the Coriolis force, and eventually forms a nuclear
ring after hitting the dust lane at the opposite side. Produced by
supersonic collisions of two gas streams, the contact points between
the ring and the dust lanes have the largest density in the ring,
and can thus be preferred sites of star formation.

Over time, the nuclear ring shrinks in size and the contact points
rotate in the azimuthal direction.\footnote{As pointed out by the
referee, Figure \ref{fig:snap} shows that the ring starts out fairly
elongated and titled relative to the bar and becomes rounder with
time. Note that the shape of the ring in Model U20 at $t=0.15\Gyr$
is remarkably similar to that of a highly-elongated nuclear ring in
ESO 565-11 observed by \citet{but99}, although the latter is
$\sim2$--$3$ times bigger than the former.} To illustrate this more
clearly, Figure \ref{fig:ring} plots the temporal and radial
variations of the azimuthally averaged surface density in the
central regions of Model U20 and its non-self-gravitating
counterpart, Model noSG. The horizontal dashed line marks the
location of the ILR. The ring is beginning to form at $t\sim0.1\Gyr$
at the galactocentric radius of $r\sim1\kpc$, well inside the ILR.
At early time when the bar is growing, the ring is not in an
equilibrium position and its shape is quite different from
$\xtwo$-orbits:  the major axis of the ring is inclined to the
$x$-axis by $-30^\circ$ at $t=0.15\Gyr$ (Fig.~\ref{fig:snap}a).  As
the ring material continuously interacts with the bar potential, the
contact points rotate in the counterclockwise direction. At
$t\sim0.2\Gyr$, the ring settles on one of the $\xtwo$-orbits, and
the contact points are located close the bar minor axis.

At the same time, the ring becomes smaller in size due to the
addition of low-angular momentum gas from outside as well as by
collisions of the ring material whose orbits are perturbed by
thermal pressure (Paper I). When self-gravity is absent, the
decreasing rate of the ring radius is $d\ln \Rring/dt \sim
-1.5\Gyr^{-1}$ until $t\simlt 0.8\Gyr$. After this time, the ring is
so small that strong centrifugal force inhibits further decay of the
ring. In Model U20 with self-gravity included, on the other hand,
strong self-gravitational potential of the ring makes the gas orbits
relatively intact, reducing the ring decay rate to $d\ln \Rring/dt
\sim -0.4\Gyr^{-1}$. The decrease of the ring size in turn causes
the contact points to move radially inward and rotate further in the
counterclockwise direction (but not more than $\sim30^\circ$).

As Figure \ref{fig:snap} shows, the bar region (i.e., inside the
outermost-$\xone$ orbit) becomes evacuated quite rapidly. This is
because the bar potential is efficient in removing angular momentum
from the gas only in the bar region, while its influence on the gas
orbits outside the outermost-$\xone$ orbit is not significant. By
$t\sim0.3\Gyr$, most of the gas in the bar region is transferred to
the ring.  The amount of the gas added to the bar region from
outside is much smaller than that lost to the ring, which causes the
mass inflow rate to the ring to decrease dramatically with time (see
Section \ref{sec:sfr}).

Figure \ref{fig:snap}d shows that at late time there is a
substantial amount of gas trapped in around an $\xone$-orbit with
$x_c=1.7\kpc$ and $y_c=4.4\kpc$, shown as a dotted line. This
elongated gaseous structure circumscribing the dust lanes and
nuclear ring is called the inner ring (e.g., \citealt{but86,reg02}):
we term the corresponding $\xone$-orbit the inner-ring
$\xone$-orbits. The formation of the inner ring in our model is as
follows. As mentioned before, the dust-lane shocks form first at far
downstream and moves toward the bar major axis as the bar potential
grows.  During this time, much of the gas in the bar region infalls
to the nuclear region. Near the time when the bar attains its full
strength, the dust lanes find their equilibrium positions on an
$\xone$-orbit, still at the leading side from the bar major axis.
The inner ring starts to form at this time, by gathering the
residual material located between the outermost and inner-ring
$\xone$-orbits. Some gases located outside the outermost
$\xone$-orbit experience collisions near the bar ends where
$\xone$-orbits crowd, and are then able to lower their orbits to the
inner-ring $\xone$-orbit, increasing the inner-ring mass. In Model
U20, the gas added to the inner ring from outside of the outermost
$\xone$-orbit is about 70\% of the total inner-ring mass at
$t=1\Gyr$. Most of the gas inside the inner-ring $\xone$-orbit had
already transited to the nuclear ring by experiencing the dust-lane
shocks before the bar was fully turned on.

\subsection{Star Formation Rate}\label{sec:sfr}

%fig4
\begin{figure*}
\hspace{0.5cm}\includegraphics[angle=0, width=17cm]{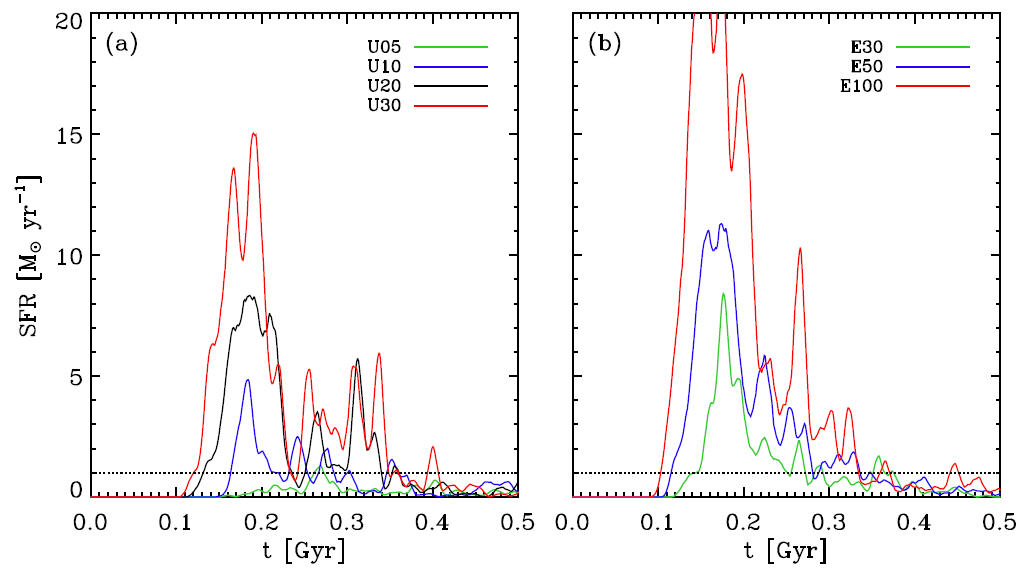}
\caption{Temporal variations, over $t=0-0.5\Gyr$, of the SFR for (a)
the uniform-disk models and (b) the exponential-disk models with
differing $\Sigma_0$. The horizontal dotted lines indicate
$\SFR=1\Aunit$. Models with larger gas mass inside the outermost
$\xone$-orbit form stars earlier and at a larger rate.
\label{fig:SFR_sig}}
\end{figure*}

%fig5
\begin{figure*}
\hspace{0.5cm}\includegraphics[angle=0, width=17cm]{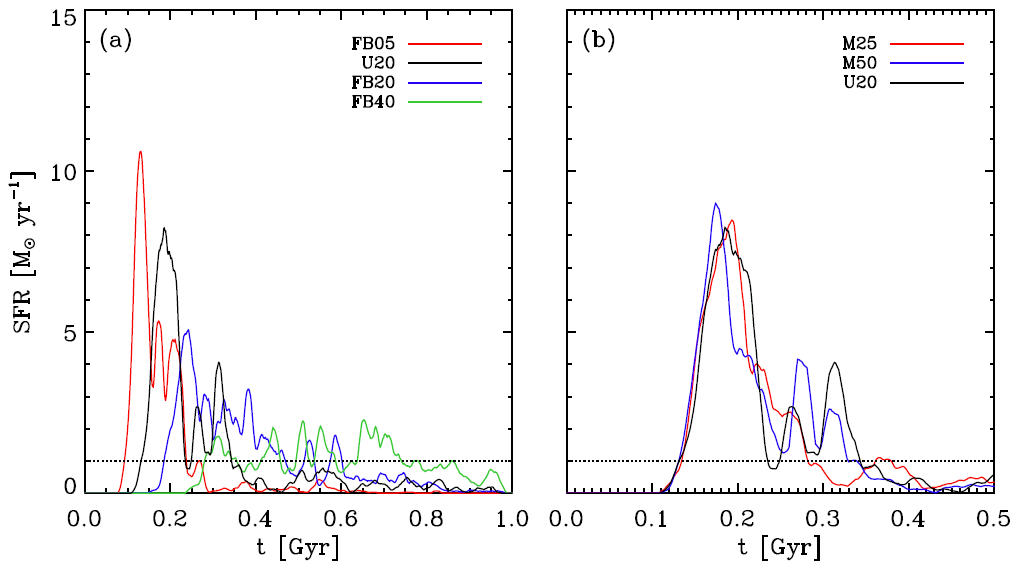}
\caption{Temporal variations of the SFR for (a) models with
differing the bar growth time $\tbar$ and (b) models with different
momentum injection $\fmom$. Note that the range of the abscissa is
$1\Gyr$ in (a) and $0.5\Gyr$ in (b). The horizontal dotted lines
indicate $\SFR=1\Aunit$. \label{fig:SFR_bar}}
\end{figure*}

Figure \ref{fig:U20} plots the time evolution of the SFR, the mass
inflow rate to the nuclear ring $\Min\equiv \int_0^{2\pi} \Sigma v_r
r d\phi$ (measured at $r=1.5\kpc$), and the total gas mass inside
the nuclear ring $\Mring$ in Model U20.  Here $v_r$ denotes the
radial velocity of the gas. In plotting these profiles, we take a
boxcar average, with a window of $20\Myr$. In this model, the bar
potential grows over the time scale of $\tbar=0.19\Gyr$, and the
first star formation takes place at the contact points at
$t=0.12\Gyr$. As the bar grows further, the dust-lane shocks become
stronger, increasing the amount of the infalling gas to the ring.
$\Min$ attains a peak value $\sim8\Aunit$ at $t=0.15\Gyr$, which
coincides with the time of highest density of the dust lanes (see
Fig.\ 5 of Paper I). The decay of $\Min$ after the peak is caused by
the fact that only the gas inside the outermost $\xone$-orbit
responds strongly to the bar potential, while the outer region is
not much affected. Similarly,  the SFR exhibits a strong burst with
a maximum value $\sim8\Aunit$, which occurs $\sim30\Myr$ after the
peak of $\Min$. The associated SN feedback produces many holes in
the gas distribution, driving a huge amount of kinetic energy to the
surrounding medium. Note that the nuclear ring, albeit somewhat
patchy, is overall well maintained despite energetic momentum
injections (Fig.~\ref{fig:snap}b).

When the mass inflow rate to the ring is very large, star formation
occurring at the contact points alone is unable to consume the whole
inflowing gas. As we will show in Section \ref{sec:azi}, the maximum
gas consumption rate afforded to the contact points is estimated to
be about $1\Aunit$ for the parameters we adopt. Any surplus
inflowing gas passes by the contact points and is subsequently added
to the ring that is clumpy. Some overdense regions in the ring are
soon able to achieve the mean density larger than the critical value
and undergo star formation, increasing the SFR rapidly. Since the
star-forming regions are randomly distributed in the ring, there is
no obvious azimuthal age gradient of star clusters in this high-SFR
phase.

Particles spawned from star formation orbit about the galaxy center
under the total gravity, but they do not feel gas pressure that is
quite strong in the nuclear ring. In addition, the gaseous ring
becomes smaller in size with time. Thus, the orbits of star
particles increasingly deviate from the gaseous orbits over time.
When stars in clusters explode as SNe, they are not always located
in the ring. A majority of star clusters are still in the ring,
while there are some clusters ($\sim10\%$) located exterior to the
ring. SNe occurring in the ring have relatively small radial
velocities due to a large background density, and the shell
expansion is limited by the surrounding dense gas.  On the other
hand, SNe exploding outside the ring can have very large expansion
velocities enough to send the neighboring gas out to the bar-end
regions. The expelled gas sweeps up the gas on its way to the
bar-end regions, passes through the dust-lane shocks again, and
falls back radially inward along the dust lane. This increases
$\Min$ temporarily during $t=0.2$--$0.35\Gyr$, with the associated
short bursts of star formation at $t=0.27$ and $0.32\Gyr$ in Model U20.

At $t=0.3\Gyr$, the bar region is almost emptied, except for the
inner ring, as most of the gas is already lost to the ring. Other
than intermittent infalls of the expelled and swept-up gas from SNe,
the mass inflow rate from the inner ring to the nuclear ring and the
related SFR become fairly small. In addition, $\Mring$ is reduced to
below $2\times10^8\Msun$ in Model U20 since star formation consumes
the gas in the ring, which also decreases the SFR (Fig.\
\ref{fig:U20}). As most of the gas in the bar region inside the
outermost $\xone$-orbit is almost lost to star formation, the galaxy
evolves into a quasi-steady state where star formation is limited to
small regions near the contact points (Fig.~\ref{fig:snap}d).

\subsection{Parametric Dependence of SFR}

%table2
\begin{deluxetable*}{c c c c c c c c c c}[t]
\tabletypesize{\footnotesize} \tablewidth{0pt}
\tablecaption{Simulation Outcomes\label{tbl:res}} \tablehead{
\colhead{Model}            & \colhead{$\mSFR$}          &
\colhead{$\dsf$}           & \colhead{$\Mtotal$}        &
\colhead{$\Mgas$}          & \colhead{$\Mring$}         &
\colhead{$\beta$}          &
\colhead{$\Gamma$}         \\
\colhead{   } & \colhead{($\rm M_\odot\;yr^{-1}$)} & \colhead{(Myr)}
& \colhead{($\rm 10^8 M_\odot$)} & \colhead{($\rm 10^8 M_\odot$)} &
\colhead{($\rm 10^8 M_\odot$)} & \colhead{   } &
\colhead{   } \\
\colhead{(1)} & \colhead{(2)} & \colhead{(3)} & \colhead{(4)} &
\colhead{(5)} & \colhead{(6)} & \colhead{(7)} & \colhead{(8)} }
\startdata
noSG&   -&  -&   -&10.66&8.03&   -&  -\\
U05 &0.53&  -& 1.4& 2.62&0.87&13.7&3.1\\
U10 &3.55& 42& 3.3& 5.33&1.39&14.4&2.6\\
U20 &7.46& 83& 9.0&10.66&1.42&12.5&2.2\\
U30 &11.8& 83&13.3&15.99&2.36&12.6&2.2\\
\hline
M25 &7.77& 87& 8.2&10.66&1.08& 9.6&2.3\\
M50 &7.78& 77& 8.7&10.66&1.55&10.8&2.2\\
\hline
FB05&9.83&68 & 8.9&10.66&1.62&12.3&2.2\\
FB20&4.04&163& 8.7&10.66&1.73& 9.7&2.2\\
FB40&1.94&392& 7.2&10.66&2.75& 7.4&3.0\\
\hline
E30 &5.10& 74& 6.1& 7.55&1.55&12.4&2.2\\
E50 &10.4& 77&10.1&12.59&1.41& 6.3&2.2\\
E100&21.9& 84&23.0&25.18&1.78& 6.3&2.0
\enddata
\tablecomments{$\mSFR$ and $\dsf$ denote the peak rate and the
duration of active star formation with $\SFR\geq \mSFR/2$,
respectively; $\Mtotal$ is the total mass in stars at $t=1\Gyr$;
$\Mgas$ is the total gas mass inside the outermost $\xone$-orbit at
$t=0$; $\Mring$ is the mass of the nuclear ring at $t=1\Gyr$; $\beta
= d\log (t/{\rm yr})/d(r/{\rm kpc})$ is the radial age gradient of
clusters; $\Gamma=-d\log N/d\log \Mstar$ is the power-law slope of
the cluster mass functions.  }
\end{deluxetable*}

%fig6
\begin{figure*}
\hspace{0.5cm}\includegraphics[angle=0, width=17cm]{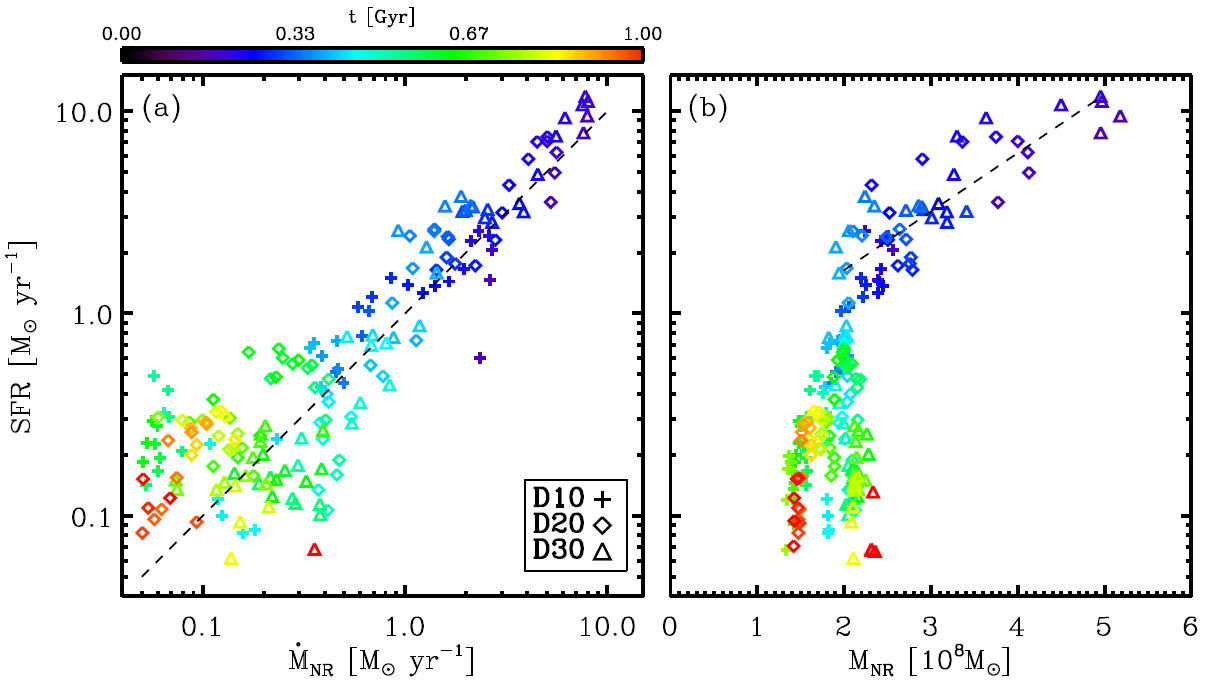}
\caption{Dependence of the SFR on (a) the mass inflow rate $\Min$ to
the ring and (b) the total mass $\Mring$ in the ring.  The dashed
line draws $\SFR=\Min$ in (a), and $\SFR\propto \Min^{0.3}$ in (b).
The SFR is almost equal to $\Min$ for the whole range of the SFR,
while it is not well correlated with $\Mring$ when
$\SFR\leq1\Aunit$. \label{fig:SFR_Min}}
\end{figure*}

Figure \ref{fig:SFR_sig} compares the SFRs from (left) uniform-disk
and (right) exponential-disk models with different $\Sigma_0$.  In
all models, the SFR displays a primary burst followed by a few
secondary bursts, with time intervals of $\sim50-80\Myr$, before
becoming reduced to below $1\Aunit$. The primary burst is associated
with the rapid gas infall due to angular momentum loss at the
dust-lane shocks, while the secondary bursts are caused by the
re-infall of the ejected gas via SN feedback out to the bar region.
Models with larger $\Sigma_0$ start to form stars earlier and have a
larger value of the maximum star formation rate, $\mSFR$. The
duration of active star formation, $\dsf$, defined by the time span when
$\rm SFR\geq \mSFR/2$, is also larger for models with larger
$\Sigma_0$, since the gas available for star formation is
correspondingly larger. Models U20 and E50 initially have a similar
gas mass inside the outermost $\xone$-orbit, but Model E50 has larger
$\mSFR$ since the gas is more centrally concentrated and thus
infalls more readily to the nuclear ring. Columns (2) and (3) of
Table \ref{tbl:res} list $\mSFR$ and $\dsf$ for all models.
The phase of active star formation lasts only for $\sim0.5\tbar$ in
all models.

%fig7
\begin{figure}
\hspace{0.5cm}\includegraphics[angle=0, width=8cm]{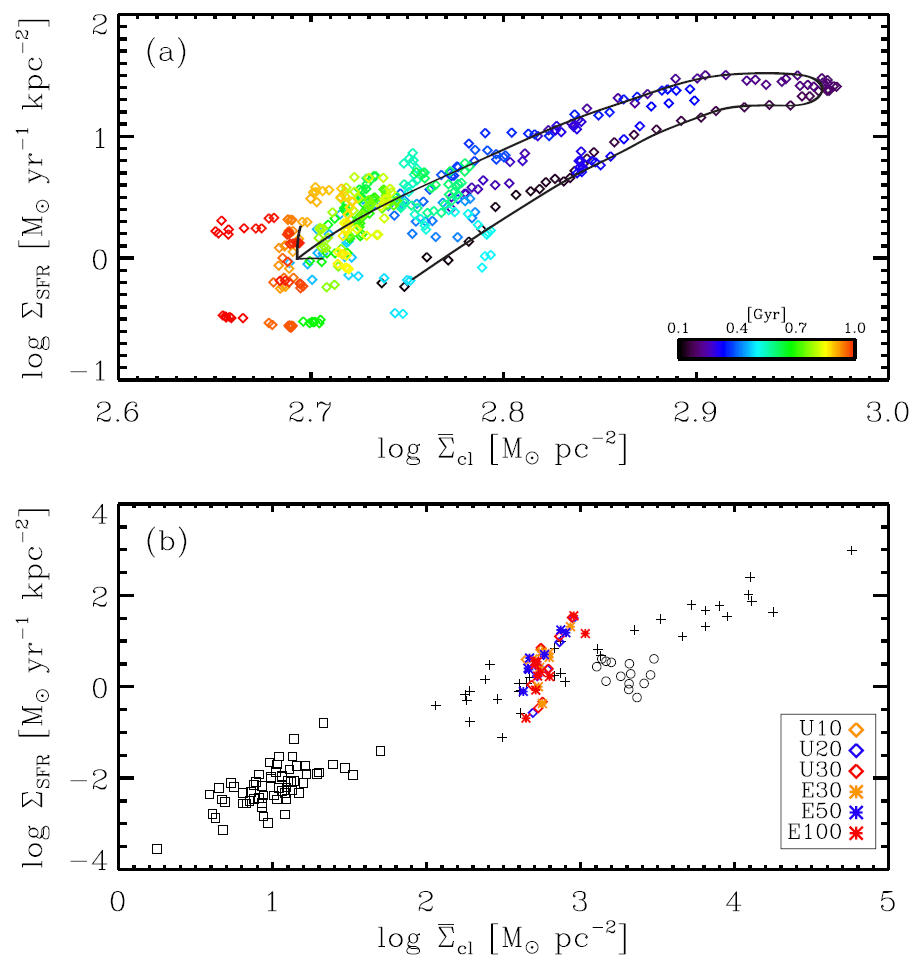}
\caption{ (a) Dependence of the SFR surface density $\SigSFR$ on the
mean surface density $\aSig$ of dense clouds for Model U20. The
colorbar indicates the epoch of star formation for each symbol, and
the curved arrow denotes the mean evolutionary track in the
$\SigSFR-\aSig$ plane. (b) $\SigSFR-\aSig$ relationship from our
models compared to the observed Kennicutt-Schmidt law. Squares and
pluses represent normal and circumnuclear starburst galaxies adopted
from \citet{ken98}, respectively, while circles are for
spatially-resolved star-forming regions in NGC 1097 from
\citet{hsi11}. \label{fig:KS}}
\end{figure}

Figure \ref{fig:SFR_bar} shows how (left) the bar growth time and
(right) the amount of the momentum injection affect the temporal
behavior of the SFR. In models where the bar grows more rapidly,
dust-lane shocks form earlier and initiate stronger gas inflows.
This causes star formation in models with smaller $\tbar$ to occur
at a higher rate and for a shorter period of time (see Table
\ref{tbl:res}). The peak SFR is attained approximately at $t\sim
(0.8-1)\tbar$. This suggests that galaxies in which the bar forms
more slowly are likely to have star formation less active
instantaneously but extended for a longer period of time. Figure
\ref{fig:SFR_bar}b shows that the SFR computed in our models is
largely insensitive to $\fmom$, although smaller $\fmom$ makes the
secondary bursts less active.

To directly address what controls star formation in nuclear rings,
we plot in Figure \ref{fig:SFR_Min} the dependence of the SFR (left)
on the mass inflow rate to the ring and (right) on the gas mass in
the ring for Models U10, U20, and U30. Color indicates the star
formation epoch of each symbol. Even though there are large scatters
especially when the SFR is low, the SFR is almost equal to $\Min$
over two orders of magnitude variations in $\Min$.  The scatters in
the $\SFR-\Min$ relation are due to the fact that star formation is
stochastic in our models and that it takes the gas some finite time
($\sim10-30\Myr$) to travel from $r=1.5\kpc$ (where $\Min$ is
measured) to the nuclear ring. On the other hand, the SFR does not
show a good correlation with $\Mring$. While $\SFR\propto
\Mring^{0.3}$ for $\SFR\simgt 1\Aunit$, it is almost independent of
$\Mring$ for $\SFR\simlt 1\Aunit$. Note that the change in $\Mring$
is less than a factor of 5 in Figure \ref{fig:SFR_Min}b, while the
SFR varies by more than two orders of magnitude. This suggests that
it is the mass inflow rate to the ring, rather than the ring mass,
that determines the SFR in the nuclear ring. Conversely, the SFR can
be a good measure of the mass inflow rate driven by the bar
potential.

Column (4) of Table \ref{tbl:res} gives the total mass in stars
$\Mtotal$ formed until the end of the run for each model.
Columns (5) and (6) list the total gas mass $\Mgas$ inside the
outermost $\xone$-orbit in the initial disk and the mass of the
nuclear ring $\Mring$ at $t=1\Gyr$, respectively. Note that $\Mgas$
is approximately the maximum gas mass available for star formation
in the ring. We find that the relation $\Mtotal = \Mgas - \Mring$,
with $\Mring=2\times 10^8\Msun$ fixed, explains the numerical
results fairly well, indicating that most of the gas inside the
outermost $\xone$-orbit flows inward to form stars, with some
residual gas remaining in the nuclear ring. Compared to Model U20
with $\tbar/\torb=1$, Model FB40 with $\tbar/\torb=4$ has $\Mtotal$
about 20\% smaller, since the gas in the bar region is still flowing
in to the nuclear ring at the end of the run.

As will be discussed in more detail in Section \ref{sec:dis}, the
overall temporal trend of the SFR (that is, rapid decline after a
primary burst except for a few secondary bursts) found in our
numerical models is largely similar to the numerical results of
previous studies (e.g., \citealt{hel94,kna95,fri95}), but appears
inconsistent with observations of \citet{all06} and \citet{sar07}
who found that star-forming nuclear rings live long, with multiple
episodes of starburst activities.  Since the ring SFR is controlled
by the mass inflows rate to the rings, this implies that rings in
real galaxies should be continually supplied with fresh gas from
outside for quite a long period of time.  Candidate mechanisms for
additional gas inflows, over a time scale much longer than the bar
growth time, include spiral arms and cosmic gas infalls, which are
not included in this paper.

\subsection{Star Formation Law}

%fig8
\begin{figure*}
\hspace{0.5cm}\includegraphics[angle=0, width=17cm]{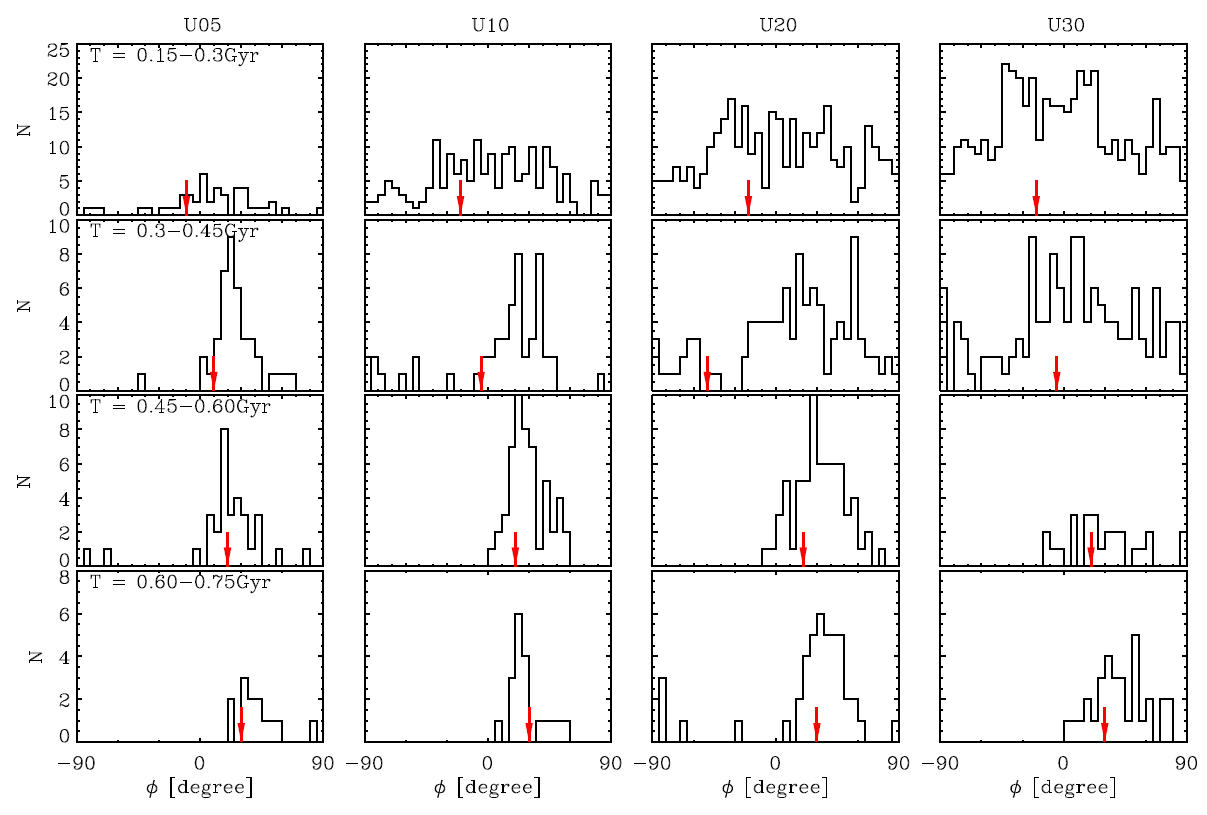}
\caption{Histograms of the star clusters that formed in each
selected time bin (with bin width of $0.15\Gyr$) for all
uniform-density models as a function of the azimuthal angle where
they form. The arrow at the bottom of each panel indicates the mean
position of a contact point. \label{fig:azh}}
\end{figure*}

To explore the dependence of the local SFR on the local gas surface
density, we define clouds as regions in the simulation domain whose
density is larger than $300\Surf$. This density roughly corresponds
to the mean density of boundaries of gravitationally bound clouds in
our models (see Section \ref{sec:clouds}). While this choice of the
minimum density for clouds is somewhat arbitrary, these clouds may
represent giant molecular clouds and their complexes including
hydrogen envelopes (e.g., \citealt{wil00,mck07}).

At each time,  we calculate the mean surface density $\aSig$ of, and
the total area $A_{\rm cl}$ occupied by, the clouds distributed
along the ring. The mean SFR surface density is then given by
$\SigSFR=\SFR/A_{\rm cl}$. Figure \ref{fig:KS}a plots the resulting
$\SigSFR$ as a function of $\aSig$ for Model U20.  Color represents
the time when each point is measured, while the curved arrow
indicates the mean evolutionary direction in the $\SigSFR$--$\aSig$
plane. When the first star formation takes place ($t=0.12\Gyr$), the ring
has $\aSig\sim560\Surf$ and $\SigSFR\sim0.5\SFRunit$. The radial gas
inflow along the dust lanes increases $\SigSFR$ rapidly until it
achieves a peak value at $t=0.18\Gyr$. The corresponding increase of
$\aSig$ is smaller since star formation reduces the gas content in
the ring. After the peak, $\SigSFR$ decreases with decreasing
$\Min$, but it has a larger value, by a factor of $\sim 2-4$ on
average, than that at the same $\aSig$ before the peak. This is
because active SN feedback stirs the ring material vigorously,
tending to increase density contrast between clumps and the
background material. After the $\SigSFR$ peak, therefore, the total
area covered by the gas with $\Sigma>300\Surf$ becomes smaller than
before, resulting in larger $\SigSFR$.

Figure \ref{fig:KS}b plots the $\SigSFR$--$\aSig$ relationship
measured at every $0.1\Gyr$ for (diamonds) the uniform-disk models
and (asterisks) the exponential-disk models. Models with larger
$\Mgas$ tend to have larger $\SigSFR$ and larger $\aSig$. Note that
our numerical results are overall consistent with the
Kennicutt-Schmidt law for normal galaxies (squares) and
circumnuclear starburst galaxies (pluses) adopted from
\citet{ken98}, and not much different from the observed
$\SigSFR$--$\aSig$ relation for spatially-resolved star-forming
regions (circles) in the nuclear ring of NGC 1097 taken from
\citet{hsi11}.\footnote{In plotting the $\SigSFR$--$\Sigma$
relation, \citet{hsi11} used the maximum density of a cloud, instead
of the mean density, for $\Sigma$.} There are large scatters in
$\SigSFR$, amounting to $\sim1$--2 orders of magnitude, both in
observational and simulation results. For star formation in nuclear
rings of barred galaxies, the gas surface density probably sets the
mean value of $\SigSFR$, as the Kennicutt-Schmidt law implies, while
the scatters in $\SigSFR$ are likely due to the temporal variations
of the mass inflow rate to the ring.

%fig9
\begin{figure*}
\hspace{0.5cm}\includegraphics[angle=0, width=17cm]{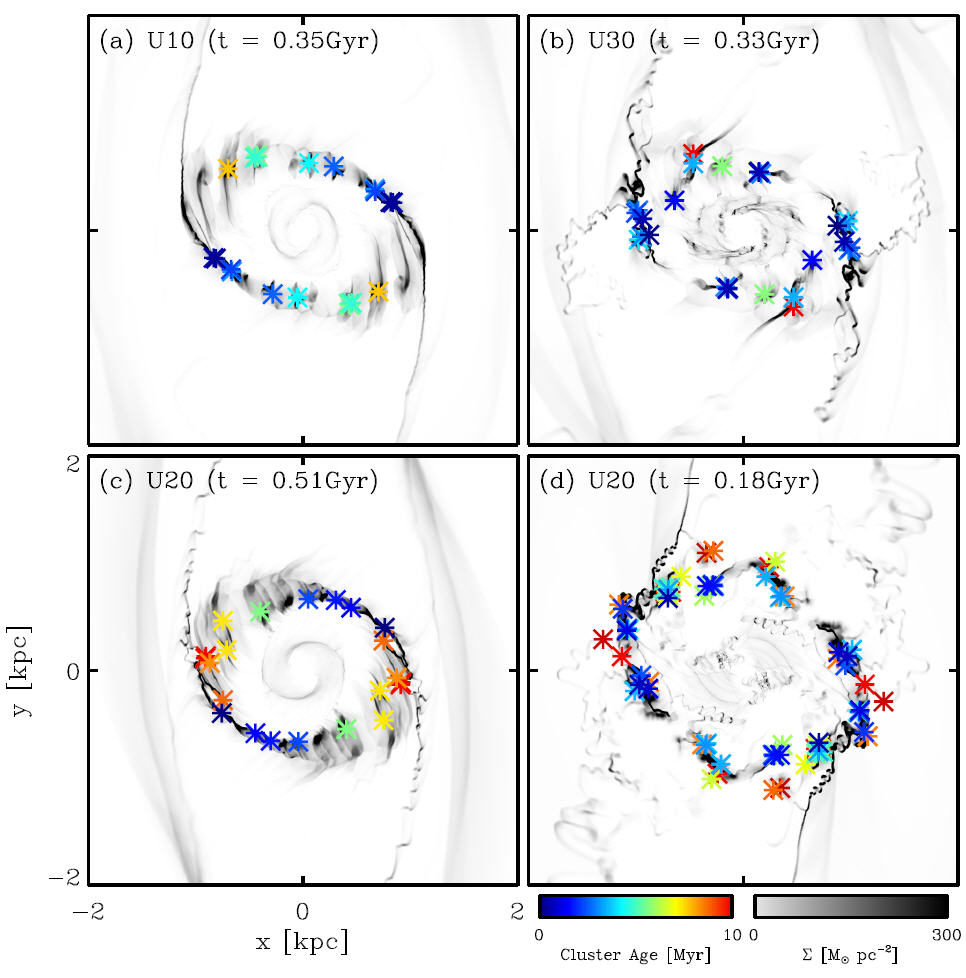}
\caption{Positions and ages of young star clusters at selected
epoches in Models U05, U20, and U30, overlaid on the gas density
distribution in linear scale. The left panels show a clear azimuthal
age gradient, while there is no age gradient in the right panels.
Color indicates the cluster ages. \label{fig:agra}}
\end{figure*}

\section{Properties of Star Clusters and Gas Clouds}\label{sec:pro}

\subsection{Azimuthal Age Gradient}\label{sec:azi}

As mentioned in Introduction, observations indicate that some
galaxies have well-defined azimuthal age gradients of star clusters
in nuclear rings (e.g., \citealt{ryd01,all06,bok08,ryd10,van13}),
while others do not (e.g., \citealt{ben02,bra12}). Our simulations
show that the presence or absence of the azimuthal age gradient is
decided by the SFR in the ring (or, more fundamentally, on $\Min$),
independent of $\Sigma_0$ and the initial gas distribution.

Figure \ref{fig:azh} plots for the uniform-density models the
histograms of star clusters formed in each selected time bin, with
bin width of $0.15\Gyr$, as a function of the angular position where
they form. The arrow at the bottom of each panel marks the location
of a contact point that is moving in the positive azimuthal
direction with time, as described in Section \ref{sec:over}. Model
U05 with $\Sigma_0=5\Aunit$ has $\SFR \simlt 0.1 \Aunit$
(Fig.~\ref{fig:SFR_sig}), and star-forming regions in this model are
almost always localized to the contact point. In Models U20 and U30,
on the other hand, star-forming regions are widely distributed along
the azimuthal direction at early time ($t=0.15-0.45\Gyr$) when $\SFR
> 1\Aunit$, while they are preferentially found near the contact
points at late time ($t\simgt 0.45\Gyr$) when $\SFR < 1\Aunit$.

Star clusters age as they orbit along the ring and emit copious UV
radiations during  about $\sim10\Myr$ after birth. If star-forming
regions are localized to the contact points, therefore, clusters
would appear as ``pearls on a string'' (e.g., \citealt{bok08}), with
an age gradient along the rotational direction of the nuclear ring.
This is illustrated in the left panels of Figure \ref{fig:agra}
which plot the spatial locations of star clusters with color
indicating their ages ($<10\Myr$), overlaid on the density
distribution in linear scale, in Model U10 at $t=0.35\Gyr$ and Model
U20 at $t=0.51\Gyr$. There is clearly a positive bi-polar age
gradient starting from the contact points that are located at
$\phi\sim30^\circ$ and $210^\circ$. On the other hand, when the
star-forming regions are randomly distributed throughout the ring,
as in the ``popcorn'' model of \citet{bok08}, star clusters with
different ages would be mixed. In this case, there is no apparent
age gradient along the ring, as exemplified in the right panels of
Figure \ref{fig:agra} for Model U30 $t=0.33\Gyr$ and Model U20 at
$t=0.18\Gyr$.

Why does then the SFR (or $\Min$) matter for the azimuthal
distributions of star-forming regions? The answer lies at the fact
there is a limit on the rate of gas consumption at the contact
points that occupy very small areas in the ring. When $\Min$ is
sufficiently small, most of the inflowing gas to the ring can be
converted into stars at the contact points, and the resulting SFR is
correspondingly small. When $\Min$ is very large, on the other hand,
the contact points cannot transform all the inflowing gas to stars
instantaneously.  The excess inflowing gas overflows the contact
points and is transferred to other regions in the ring.  The ring
becomes denser not only by the addition of the overflowing gas but
also by its own self-gravity.  Some clumps in the ring achieve
surface density above the threshold value, and start to form stars.
We find that the rings have the Toomre stability parameter as low as
$\sim0.5$ when the SFR is near its peak, suggesting that
gravitational instability promotes star formation.

%fig10
\begin{figure*}
\hspace{0.5cm}\includegraphics[angle=0, width=17cm]{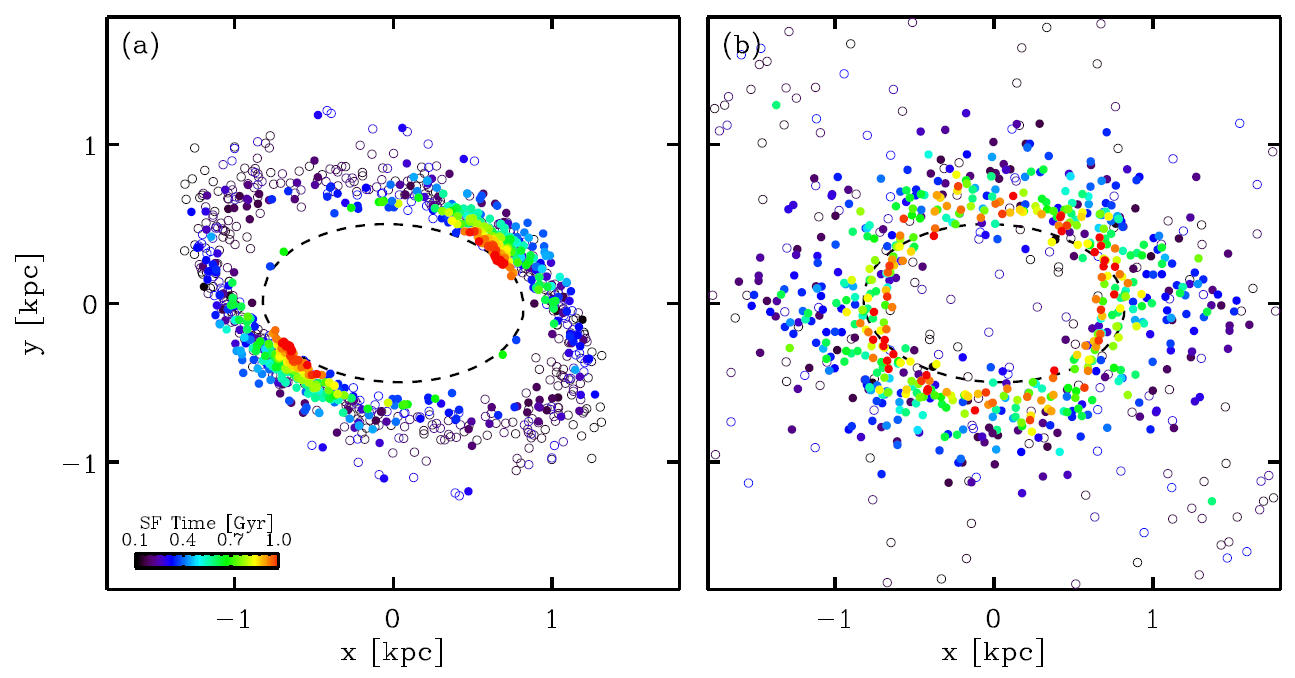}
\caption{Spatial distributions of (a) the formation locations and
(b) the present positions of star clusters in Model U20 at
$t=1\Gyr$. The dashed lines draw the ring at this time. Open circles
denote the clusters that have passed the central region with
$r<0.3\kpc$ during their orbits, while filled circles represent
those that have not. Colorbar indicates the formation epoch of the
clusters in unit of Gyr.  At late time, stars form preferentially
near the contact points. Gravitational interactions lead to
diffusion of the clusters. \label{fig:radpos}}
\end{figure*}

Observations show that galaxies with no azimuthal age gradient
have, on average, slightly larger SFRs in the rings \citep{maz08},
although the spread in SFR for individual galaxies is too large to
make this conclusive. Note, however, that the critical SFR
determining the presence or absence of the azimuthal age gradient
depends on many parameters. The maximum SFR at the contact points
can be estimated as follows. Let $\Delta r$ and $\Delta \phi$ denote
the radial thickness and the azimuthal extent of a contact point,
respectively. Then, the maximum SFR expected from two contact points
is simply
\begin{equation}\label{eq:sfcp}
\mcp = 2\epsilon_{\rm ff} \Sigma_{\rm CP} \Rring
\Delta r\Delta \phi/t_{\rm ff},
\end{equation}
where $\Sigma_{\rm CP}$ is the surface density of the contact
points. The mean density of star-forming clouds is $\sim4000\Surf$,
which we take for $\Sigma_{\rm CP}$.  For $\epsilon_{\rm ff}=0.01$,
$\Delta r=50\pc$, $\Delta \phi=30^\circ$, and $\Rring=1\kpc$ typical
in our models, equation (\ref{eq:sfcp}) yields $\mcp \sim 1\Aunit$,
consistent with our numerical results.  Note that the specific value
of $\mcp$ depends on the parameters we adopt.  In particular, $\mcp
\propto c_s^3\Rring^2/\rSF^{3/2}$ if the ring width is proportional
to the ring radius, suggesting that galaxies with a weak bar (to
have a smaller ring) and strong turbulence would have large $\mcp$.

\subsection{Radial Age Gradient}

Although the presence of an azimuthal age gradient depends on the
SFR, we find that star clusters always display a radial age
gradient. Figure \ref{fig:radpos} plots the spatial distributions of
(left) the formation locations and (right) the present positions of
star clusters on the $x$-$y$ plane at $t=1\Gyr$ in Model U20. Each
cluster is colored according to its formation time. The open circles
denote the clusters that have passed the central region with
$r<0.3\kpc$ at least once during their orbital motions: such
clusters would have been destroyed at least partially by strong
tidal fields near the galaxy center if their internal evolution such
as core collapse, evaporation, disruption, etc.\ had been
considered. On the other hand, the filled circles are for clusters
that have never approached the central region, and thus are most
likely to survive the galactic tide. The dashed lines draw the ring
at $t=1\Gyr$. Star-forming regions at late time are concentrated on
the contact points, while they are well distributed at early time.
Note that clusters that form early before the nuclear ring settles
on an $\xtwo$-orbit  have initial kick velocities quite different
from those on $\xone$- or $\xtwo$-orbits at their formation
locations.  Although the ring soon takes on the $\xtwo$-orbit, these
clusters move on eccentric orbits and wander around the nuclear
region. Figure \ref{fig:radpos}b shows that young clusters are
preferentially found near the ring, while old clusters are located
away from it, indicative of a positive radial gradient of their
ages.

%fig11
\begin{figure}
\hspace{0.5cm}\includegraphics[angle=0, width=8cm]{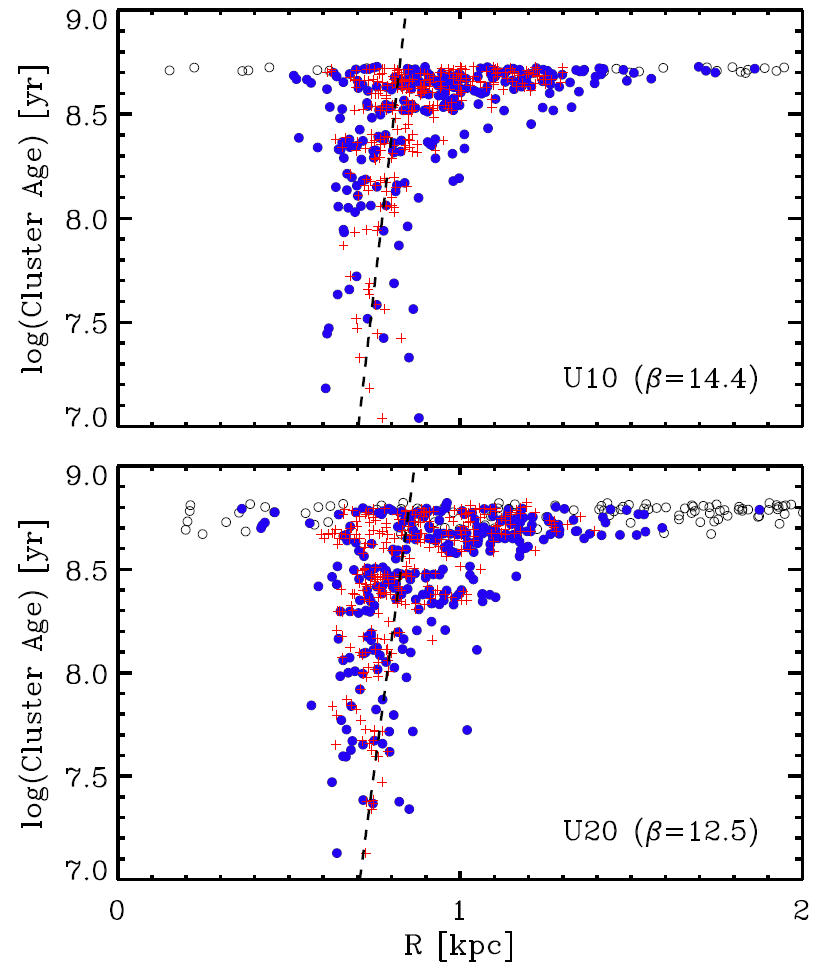}
\caption{ Ages of star clusters as functions of their current radial
locations (circles) at $t=1\Gyr$ and their formation positions
(pluses) for Models U10 and U20. Open circles are those that have
passed by the galaxy center at a very close distance during orbital
motions, while the filled circles are for those that have not. The
dashed lines are the fits, with slopes of $\beta= d\log (t/{\rm
yr})/d(r/{\rm kpc}) =14.4$ and $12.5$, for the initial cluster
positions, for Models U10 and U20, respectively. \label{fig:radgra}}
\end{figure}

To show this more clearly, Figure \ref{fig:radgra} plots the age of
clusters as functions of their present radial positions (circles) at
$t=1\Gyr$ as well as their formation locations (plus symbols) for
Models U10 and U20.  Again, the open circles denote the clusters
that have passed by the galaxy center, while the filled circles are
for those that have not.  Note that the age distributions of the
present-day and formation-epoch locations of the clusters are not
much different from each other, although the former shows a large
spatial dispersion. The dispersion is larger for older clusters. To
quantify the radial age gradient, we bin the clusters according to
their ages, with a bin size of $\Delta \log (t/{\rm yr}) = 0.2$, and
calculate the mean age and position in each bin. Our best fits of
the ages to the formation-epoch positions are $\beta \equiv d\log
(t/{\rm yr})/d(r/{\rm kpc})\sim 14.4$ and $12.5$ for Models U10 and
U20, respectively. Column (7) of Table \ref{tbl:res} gives $\beta$
for all models. This radial age gradient results primarily from the
decrease in the ring size with time, such that old clusters formed
at larger galactocentric radii. Clusters diffuse out radially
through gravitational interactions themselves and also with dense
clouds in the ring, without much effect on the radial age gradient.

%fig12
\begin{figure*}
\hspace{0.5cm}\includegraphics[angle=0, width=17cm]{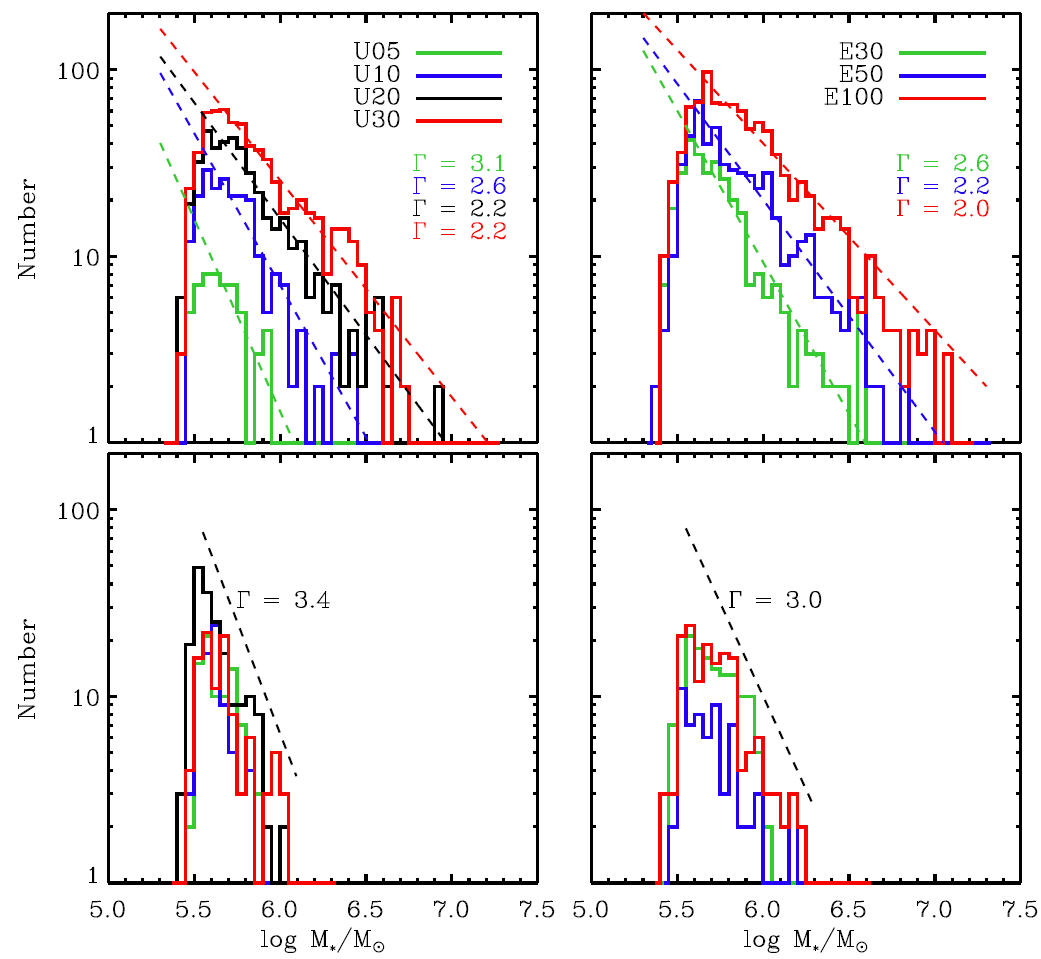}
\caption{Clusters mass functions for (left) the uniform-density and
(right) the exponential-density models. The upper and lower panels
plot the clusters formed when the SFR is larger or smaller than
$1\Aunit$, respectively.  The mass function becomes shallower with
increasing SFR. \label{fig:mf}}
\end{figure*}

\subsection{Cluster Mass Functions}

Figure \ref{fig:mf} plots the mass functions of all the clusters
that have formed in each of the (left) uniform-disk and (right)
exponential-disk models until $t=1\Gyr$.  The upper panels are for
the clusters formed while star formation is very active with
$\SFR\geq 1\Aunit$, whereas those produced when $\SFR < 1\Aunit$ are
presented in the lower panels. In general, the mass distribution of
clusters is described roughly by a power law, with its index
depending on the SFR and $\Mgas$. When the SFR is larger than
$1\Aunit$, the slope of the mass function is $\Gamma\equiv - d\log
N/d\log \Mstar \sim2$--$3$, with a smaller value corresponding to
larger $\Mgas$. When $\SFR\leq1\Aunit$, clusters have a much steeper
mass distribution with $\Gamma\simgt 3$. Column (8) of Table
\ref{tbl:res} gives $\Gamma$.

The dependence of the power-law index on the SFR and $\Mgas$ can be
understood as follows. When the SFR is large, there are numerous
dense regions distributed throughout the ring. Such regions grow by
accreting the surrounding material. Since star formation occurs in a
stochastic manner in our models, some dense clouds have a chance to
grow as massive as, or even larger than $\sim10^7\Msun$, leading to
a relatively shallow mass function. On the other hand, when the SFR
is small, there are not many dense regions.  Since the growth of
density is quite slow in these models, they form stars at densities
slightly above the critical value.  In this case, most clusters have
mass around $\sim10^5-10^6\Msun$, with a steep mass distribution.

\subsection{Giant Clouds}\label{sec:clouds}

Finally, we present the properties of giant clouds located in
nuclear rings.  High-resolution radio observations show that nuclear
rings consist of giant molecular associations at scale of
$\sim0.2$--$0.3\kpc$ in which most star formation takes place. They
typically have masses of $\sim10^7\Msun$ and are gravitationally
bound (e.g., \citealt{hsi11}).

\begin{figure*}
\hspace{0.5cm}\includegraphics[angle=0, width=17cm]{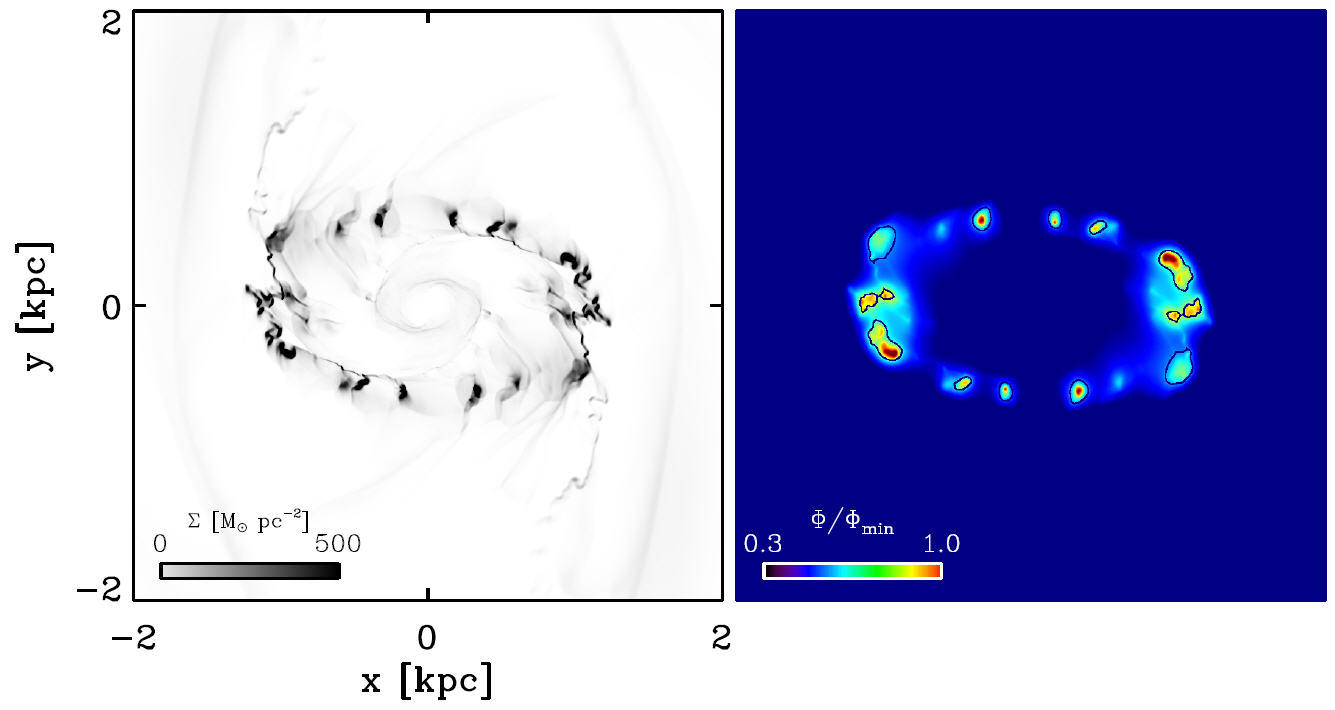}
\caption{Distribution of giant clouds in the nuclear ring of Model
U20 at $t=0.36\Gyr$.  Left: the gas surface density is shown in
linear scale. Right: cloud boundaries found by the method described
in the text are overlaid on the gravitational potential of the gas.
\label{fig:clump}}
\end{figure*}

To identify giant clouds in our models, we utilize a core-finding
technique developed by \citet{gon11}.  This method makes use of the
gravitational potential of the gas, and thus allows smoother cloud
boundaries than the methods based on isodensity surfaces (see, e.g.,
\citealt{smi09}). At a given time, we search for all the local
minima of the gravitational potential and find the largest closed
potential contour encompassing one and only one potential minimum.
We then define the potential minimum and outermost contour as the
center and boundary of a cloud, respectively. If the distance
between two neighboring minima is less than $0.1\kpc$, we combine
them. Figure \ref{fig:clump} plots, for example, giant clouds
identified by this technique for Model U20 at $t=0.36\Gyr$.  The
left panel shows the gas surface density in linear scale, while the
right panel displays boundaries of giant clouds as contours overlaid
over the gravitational potential of the gas. A total of 14 giant
clouds are identified.  The mean values of their masses $M$, radii
$R$, and one-dimensional velocity dispersions $\sigma$ are
$10^7\Msun$, $100\pc$, and $20\kms$, respectively, corresponding to
supersonic internal motions. The average density of the cloud
boundaries is found to be $\sim300\Surf$. The average value of the
virial parameter is $\alpha = 5\sigma^2R/(GM) \sim 2$, so that they
are gravitationally bound, consistent with the observed cloud
properties in nuclear rings of barred galaxies (e.g.,
\citealt{mck07}).

%fig13

\section{Summary and Discussion}

\subsection{Summary} \label{sec:sum}

We have presented the results of two-dimensional hydrodynamic
simulations of star formation occurring in nuclear rings of barred
galaxies. We initially consider an infinitesimally thin, isothermal
gas disk placed under the external gravitational potential.  The
external potential consists of a stellar disk, a stellar bulge, a
central BH, and a non-axisymmetric stellar bar.  We do not study the
effect of spiral arms in the present work. The bar potential is
modeled by a Ferrers prolate spheroid with the semi-major and minor
axes of $5\kpc$ and $2\kpc$, respectively, and rotates about the
galaxy center with a patten speed of $33\freq$. The bar mass, when
it is fully turned on, is set to 30\% of the total stellar mass in
the spheroidal component, corresponding to a strongly barred galaxy.
We fix the gas sound speed to $\cs=10\kms$ and the BH mass to
$4\times10^7\Msun$.  Our simulations incorporate star formation
recipes that include a density threshold corresponding to the Jeans
condition, a star formation efficiency, conversion of gas to
particles representing star clusters or their groups, and delayed
momentum feedback via SN explosions. To explore various situations,
we consider both uniform and exponential density models, and vary
the gas surface density, bar growth time, and the total momentum
injection in the in-plane direction. The main results of this work
can be summarized as follows.

The imposed bar potential readily induces a pair of dust-lane shocks
in the bar region inside the outermost $\xone$-orbit. At early time
when the bar potential is weak, the dust-lane shocks are placed at
the far downstream side from the bar major axis.  As the bar
potential increases, the dust-lane shocks become stronger and slowly
move toward the bar major axis. The gas passing through the shocks
loses a significant amount of angular momentum, infalls radially
along the dust lanes, and forms a nuclear ring. The continuous gas
inflows provide a fuel for star formation in the ring. After the bar
potential reaches its full strength, the dust lanes settle on an
$\xone$-orbit, while the nuclear ring follows an $\xtwo$-orbit.  The
remaining gas located inside the outermost $\xone$-orbit and outside
the dust lanes is gathered to form an elongated inner ring in the
bar region, whose shape is well described by an $\xone$-orbit, as
well. Some of the gas located outside the outermost $\xone$-orbit
transits to the inner ring near the bar ends where $\xone$-orbits
crowd.  Similarly, the gas in the inner ring loses angular momentum
when it collides with other gas near the bar ends, slowly
infalling to the nuclear ring through the dust lanes.

The contact points between the dust lanes and the nuclear ring is a
tunnel through which the inflowing gas on $\xone$-orbits switches to
the $\xtwo$-orbit of the nuclear ring. About the time when the bar
potential is fully turned on, the contact points are located near
the bar minor axis. Over time, the nuclear ring shrinks in size due
to the addition of low angular momentum gas from outside and by
collisions of the ring material, which in turn makes the contact
points rotate slowly in the counterclockwise direction. Since the
contact points have largest density in the nuclear ring, they are
preferred sites of star formation, although star-forming regions can
be distributed throughout the ring when the mass inflow rate is
high.

The bar potential transports the gas in the bar region to the
nuclear ring very efficiently, but does not have strong influence on
the gas orbits outside the bar region. This not only makes the bar
region evacuated rapidly but also reduces the mass inflow rate
$\Min$ dramatically after $\sim 0.3\Gyr$. The SFR in nuclear rings
displays a single primary burst followed by a few secondary bursts
before becoming reduced to small values. The primary burst is
associated with the massive gas inflow along the dust lanes caused
by the growth of the bar potential. The duration and maximum rate of
the primary burst depend on the growth time of the bar potential, in
such a way that a slower bar growth results in a more prolonged and
reduced SFR. The secondary bursts are due to the re-entry of the
ejected and swept-up gas by SN feedback from the nuclear ring out to
the bar-end regions.  Time intervals between the secondary bursts
are roughly  $\sim50-80\Myr$. The peak SFR is attained at
$t\sim(0.8-1)\tbar$, and the duration of active star formation is
roughly a half of the bar growth time. The SFR is almost equal to
$\Min$. It has a weak dependence on the total gas mass $\Mring$ in
the ring when $\SFR\simgt1\Aunit$, and is not correlated with
$\Mring$ when $\SFR\simlt1\Aunit$. This suggests that star formation
in the ring is controlled primarily by $\Min$ rather than $\Mring$.
The relationship between the SFR surface density and the surface
density of dense clumps in nuclear rings found from our numerical
models are consistent with the usual Kennicutt-Schmidt law for
circumnuclear starburst galaxies.

The presence or absence of azimuthal age gradients of young star
clusters in nuclear rings depends on the SFR (or $\Min$) in our models.
When $\Min$ is small, most of the inflowing gas to the nuclear ring
is consumed at the contact points.  In this case, young star
clusters that form would exhibit a well-defined azimuthal age gradient
along the ring. When $\Min$ is large, on the other hand, the contact
points are unable to transform all of the inflowing gas to stars.
The extra gas overflows the contact points and goes into the nuclear
ring.  The ring becomes massive and forms stars in clumps that become
dense enough. In this case, no apparent age gradient of star
clusters is expected since star-forming regions are randomly
distributed over the whole length of the ring. The critical value of
$\Min$ that determines the presence or absence of the azimuthal age
gradient is estimated to be $\sim 1\Aunit$ in our models,
although it depends on various parameters such as the ring radius,
critical density, etc.\ (eq.~[\ref{eq:sfcp}]).

Star clusters produced also exhibit a positive radial age gradient,
such that young clusters are located close to the nuclear ring,
while old clusters are found away from the ring.  The primary reason
for this is that the nuclear ring becomes smaller in size with time,
and thus star-forming regions gradually move radially inward. In our
models, the radial age gradient amounts to $\beta=d\log (t/{\rm
yr})/d(r/{\rm kpc}) \sim 6-15$. Radial diffusion of clusters via
mutual gravitational interactions and also with the gaseous ring
does not affect the radial age gradient much.

When the SFR is large ($>1\Aunit$), some dense clouds are able to grow
by accreting surrounding material and form massive clusters. The
cluster mass function is well described by a power law, with slope
$\Gamma=-d\log N/d\log \Mstar \sim 2-3$. A larger slope corresponds
to a more massive disk in which more gas is available for the ring
star formation. When the SFR is small, on the other hand, most clusters
form near the threshold density, leading to a steeper slope with
$\Gamma\simgt 3$. Giant clouds in nuclear rings have typical masses
$10^7\Msun$ and sizes $0.1\kpc$.  Driven by momentum injection from
SNe, their one-dimensional internal velocity dispersions are
supersonic at $\sim 20\kms$. They are gravitationally bound with the
virial parameter of $\alpha\sim2$.

\subsection{Discussion} \label{sec:dis}

We find that the SFR in nuclear rings shows a strong primary burst,
with its duration and peak value dependent on the bar growth time,
and subsequently a few weak and narrow bursts, after which the SFR
becomes very small. The peak of the primary burst is attained
roughly when the bar potential is fully turned on. This burst
behavior of the SFR appears to be a generic feature of star
formation in nuclear rings of strongly-barred galaxies found in
numerical simulations. For instance,  $N$-body$+$ SPH models
presented by \citet{hel94}, \citet{kna95}, and \citet{fri95} showed
that the SFR reaches its peak value, with narrow bursts
superimposed, about the time when the stellar bar fully develops,
after which it is reduced to small values.
In numerical modeling for star formation in the nuclear region of
the Milky Way, \citet{ksj11} found that the SFR is maximized at
$t\sim0.15\Gyr$, with a peak value $\sim0.22\Aunit$, and then drops
to a relatively constant value $\sim0.05-0.07\Aunit$. The sustained
star formation in \citet{ksj11} is thought to arise because the
Milky Way has a very weak bar that takes a long time to clear out
gas in the bar region. In this case, the gas infall may proceed
continuously over an extended period of time, a situation similar to
the case with a slowly-growing bar.

There is observational evidence that star formation in nuclear rings
occurs continually over a long period of time (a few Gyrs) with
successive $\sim4-10$ bursts separated by a few tenths of Gyrs each
(e.g., \citealt{all06,sar07}; see also \citealt{van13}). This is in
sharp contrast to our numerical results that show that star
formation in nuclear rings is dominated by one primary burst before
declining to small values, with $\sim 0.1\Gyr$ duration of active
star formation. This discrepancy is mostly likely due to the fact
that our models consider only a bar potential for angular momentum
transport and thus are too simple to describe more complicated mass
inflows in real disk galaxies. If the mass inflow rate to a nuclear
ring really controls the SFR in the ring, as found by our numerical
models, the observational results call for a need to feed the rings
with gas episodically for a long time interval. \emph{The bar
potential alone is unlikely responsible for gas supply needed for
star formation in real nuclear rings.} Unless bars are dynamically
young, present star formation in nuclear rings of nearby barred
galaxies requires additional gas feeding. One obvious such mechanism
is spiral arms that can remove angular momentum at spiral shocks to
transport gas from outer disks to the bar regions (e.g.,
\citealt{lub86,ks12}), which is not considered in the present work.
Accretion of halo gas to the disk may not only rejuvenate bars
(e.g., \citealt{bou02}) but also enhance the SFR in the rings (e.g.,
\citealt{jia99,fra06,fra08}). Such gas flows might actually exist,
as evidenced by the presence of an enhanced number of carbon stars
in the outer spiral arms of M33 \citep{blo07}. Temporal variations
in the bar strength (e.g., \citealt{bou02}) and in the bar pattern
speed (e..g, \citealt{com81}) are also likely to affect the mass
inflow rate to the ring and thus the SFR.

Some galaxies such as IC 4933 \citep{ryd10} show age gradients of
star clusters along the azimuthal direction in nuclear rings, while
there are other galaxies such as NGC 7552 \citep{bra12} that do not
show a clear age gradient. \citet{maz08} analyzed H$\alpha$ data for
22 nuclear rings and found that about half of their sample galaxies
contain azimuthal age gradients, although most of them are not
throughout the entire ring. They also found that the mean SFR in
galaxies with azimuthal age gradients is $2.2\pm0.7\Aunit$, which is
slightly larger than the mean value of $3.6\pm1.1\Aunit$ for
galaxies with no apparent age gradient.  While this appears
consistent with our numerical results, the large dispersions in the
mean SFRs suggest that there is no fixed SFR that can distinguish
between galaxies with and without age gradients. In addition, the
critical SFR for the absence or presence of azimuthal age gradients
is about $1\Aunit$ in our models, while most galaxies in the sample
of \citet{maz08} have $\SFR>1\Aunit$. As equation (\ref{eq:sfcp})
suggests for the maximum SFR, $\mcp$, allowed at the contact points,
however, the critical SFR depends on many factors that may vary from
galaxy to galaxy. For example, NGC 1343 with the most clear bi-polar
age gradient in the \citet{maz08} sample has the current SFR of
$\sim6.8\Aunit$.  Its ring radius is $\sim2\kpc$ \citep{com10},
which increases the critical SFR by a factor of 4, assuming that the
ring width is proportional to the ring size and the other parameters
remain the same.  In addition, $\mcp \propto \Sigma_{\rm
CP}^{3/2}\propto c_s^3/\rSF^{3/2}$, so that the level of
interstellar turbulence and the size of star-forming regions $\rSF$
may change $\mcp$ considerably.

We find that star clusters that form in nuclear rings naturally
develop a positive radial gradient of their ages owing primarily to
the decrease in the ring size over $\sim\Gyr$ in our models. This is
consistent with the results of \citet{jan13} who found that clusters
with ages $\simlt 1\Gyr$ in the nuclear region of NGC 1672 are older
systematically with increasing radius.  Note that the radial age
gradient holds over a timescale of $\sim\Gyr$ and may not apply to
clusters in a small age range since the decay of the ring size is
quite slow. Indeed, \citet{maz08} found that two (NGC 5953 and 7570)
of their sample galaxies show a negative radial age gradient of
\ion{H}{2} regions in the rings.\footnote{Note that NGC 5953 is a
non-barred galaxy.} Our numerical results plotted in Figure
\ref{fig:radgra} also show that when limited to clusters with age
$\sim 10^7$--$10^{7.5}$ yr at $t=1\Gyr$, younger clusters can be
found at larger radii, which is due to the stochastic nature of star
formation and ensuing gravitational interactions.

Our results show that the SFR in the nuclear rings is tightly
correlated with the mass inflow rate to the ring rather than the
total gas mass in the ring (Fig.~\ref{fig:SFR_Min}). This result is
seemingly consistent with the results of \citet{ben02} who found
that the SFR in the nuclear ring of a strongly-barred galaxy NGC
4314 is smaller, by a factor of 30, than that in a weakly-barred
galaxy NGC 1326 \citep{but00}, even if the gas mass in the ring is
smaller by only a factor of two. It is interesting to note that the
gas mass contained in most ``gas-rich'' nuclear rings of barred
galaxies in the BIMA SONG sample is in a remarkably narrow range of
$\sim(1-6)\times 10^8\Msun$ \citep{she05}.\footnote{While
\citet{she05} reported that the ring in NGC 6946 has a mass of
$\sim10^9\Msun$, a higher-resolution observation of \citet{sch06}
gives the ring mass of $\sim4\times10^8\Msun$.}  The SFR data
presented in \citet{maz08} combined with the bar strength given in
\citet{com10} show that strongly-barred galaxies usually have very
small present-day SFRs and the SFRs in weakly-barred galaxies vary
in a wide range, although the number of galaxies in their sample is
too limited to make a conclusive statement. It will be interesting to
see how the bar strength as well as gas inflows by spiral shocks
influence the SFR in nuclear rings.

While star formation is concentrated in nuclear rings in our models,
observations indicate that star formation in some galaxies occurs
not only in nuclear rings but also in the bar region including dust
lanes  (e.g., \citealt{mar97,she00,zur08,elm09, mar11}).  While dust
lanes themselves are known hostile to star formation due to strong
velocity shear (e.g., \citealt{ath92,kim12}), \citet{she00} proposed
that stars form in interbar dust spurs in filamentary shape that
impact the dust lanes from the trailing side of the bar (see also
\citealt{she02,zur08}). Indeed, \citet{elm09} inferred that some
clusters in the nuclear ring of NGC 1365 actually formed in one of
the dust lanes by the impact of spurs and subsequently migrated
inward to the nuclear ring. The origin of these interbar spurs is
yet unclear. Apparently, there is no filamentary interbar feature in
our models. They may originate from gas inflows due to spiral shocks
from the region outside the bar \citep{elm09},  from interactions of
gas with magnetic fields that are pervasive in the bar region
\citep{bec99,bec05}, and/or from other dynamical processes that
involve gas cooling, self-gravity, etc., which are not considered in
the present work.  It will be an important direction of future work
to study how spiral arms and magnetic fields affect the gas inflows
and star formation in the bar and nuclear regions.

\acknowledgments We gratefully acknowledge I.~S.\ Jang and M.~G.\
Lee for sharing their results on the radial age gradient of clusters
found in NGC 1672.  We also thank E.~C.\ Ostriker and K.\ Sheth for
helpful discussions, and are grateful to the referees for an
insightful report and for the information on ESO 565-11. This work
was supported by the National Research Foundation of Korea (NRF)
grant funded by the Korean government (MEST), No.\ 2010-0000712. The
computation of this work was supported by the Supercomputing
Center/Korea Institute of Science and Technology Information with
supercomputing resources including technical support
(KSC-2012-C3-19).

\end{document}